\documentclass[12pt,preprint]{aastex}
\shorttitle{Intermediate-redshift DLAs}
\shortauthors{Khare et al.}

\begin{document}


\title{Metals and Dust in Intermediate-redshift Damped Ly-$\alpha$ 
Galaxies}


\author{Pushpa Khare\altaffilmark{1}}
\affil{Dept. of Physics, Utkal University, Bhubaneswar, 751004, India}

\author{Varsha P. Kulkarni}
\affil{Dept. of Physics and Astronomy, Univ. of South Carolina, 
Columbia, SC 29208}

\author{James T. Lauroesch}
\affil{Dept. of Astronomy, Northwestern University, Evanston, IL, 60208}

\author{Donald G. York\altaffilmark{2}}
\affil{Dept. of Astronomy and Astrophysics, University of Chicago, 
Chicago, IL 60637}

\author{Arlin P. S. Crotts}
\affil{Dept. of Astronomy, Columbia University, New York, NY, 10027}

\author{Osamu Nakamura}
\affil{School of Physics and Astronomy, University of Nottingham,
Nottingham, NG7 2RD, UK}


\altaffiltext{1}{Also, Dept. of Physics and Astronomy, Univ. of South Carolina, 
Columbia, SC 29208 }
\altaffiltext{2}{Also, Enrico Fermi Institute, University of Chicago, Chicago, IL 60637.}


\begin{abstract}

We report spectroscopic observations with the Multiple Mirror Telescope
for 11 damped Lyman-alpha absorbers (DLAs) or strong DLA candidates at
$0.1 < z < 1.5$, including several absorbers discovered in the Sloan
Digital Sky Survey. In particular, we have measured absorption lines of
Zn II, Cr II, Ni II, Fe II, Mn II, Ti II, Ca II, and Si II. These
measurements have doubled the sample of Zn and Cr measurements at $z <
1$. The average relative abundance patterns in these objects are very
similar to those found for high-redshift DLAs reported in the literature.
Our observations suggest that the dust content, as determined by [Cr/Zn],
does not show much change with redshift. We also examine the sample for
correlation of [Cr/Zn] with estimates of the quasar reddening. Our data
suggest that the global mean metallicity of DLAs, as measured by the gas
phase abundance of Zn, at best shows a weak evolution with redshift over
the range $0.4 < z <3.9$. \end{abstract}


\keywords{quasars: absorption lines; galaxies: evolution; 
galaxies: abundances; cosmology: observations}


\section{Introduction}

Observations of metallicity evolution in galaxies provide important
constraints on the histories of gas consumption and star formation,
because the global densities of gas, metals, and stars are closely
coupled. Most cosmic chemical evolution models (e.g., Pei \& Fall, 1995;
Malaney \& Chaboyer 1996; Pei, Fall, \& Hauser 1999; Tissera et al. 2001;
Somerville, Primack \& Faber  2001)  predict that the global mean
interstellar metallicity in galaxies rises  from nearly zero at high
redshifts to nearly solar at $z=0$. The present-day mass-weighted mean
metallicity of local galaxies is also nearly solar (see, e.g., Kulkarni
\& Fall 2002). It is of great importance to determine whether the global
mean metallicities of galaxies at intermediate and high redshifts agree
with predictions of the cosmic chemical evolution models. 

There is a long history of using damped Lyman-alpha absorbers (DLAS) to
measure the gas phase abundances over the age of the Universe. It has
been believed that DLAs constitute most of the neutral gas in the
galaxies at high redshifts, enough to form all stars visible today (Wolfe
et al. 1995). Furthermore, DLAs were also believed to provide an unbiased
sample of normal galaxies as they are selected only through the presence
of large amounts of neutral hydrogen. Recent data have raised questions
regarding the validity of both these assumptions (Boisse et al. 1998; Rao
\& Turnshek, 2000; Kulkarni et al. 2001; Rao et al., 2003). Presence of
dust and gravitational lensing can also affect the DLA statistics (Fall
\& Pei, 1993; Bartelman \& Loeb, 1996). However, DLAs are still the
primary source of information about the chemical abundances in galaxies
over $ > 90 \%$ of the age of the Universe. A large number of elements
have been observed in DLAs. In particular, Fe, Zn, Si, S and O have been
used as probes of the metallicity in these systems. We prefer to use Zn to
determine the total (gas + solid phase) metallicity in DLAs. This is
because (i) Zn tracks Fe in most Galactic stars (e.g., Mishenina et al.
2002); (ii) it is relatively undepleted on interstellar dust grains; and
(iii) the lines of the dominant ionization species Zn~II are most often
unsaturated. Abundances of depleted elements such as Cr, Fe or Ni
relative to Zn probe the dust content of the absorbers (e.g., Pettini et
al. 1997; Kulkarni, Fall, \& Truran 1997).  

DLAs are thus useful for testing predictions of cosmic chemical evolution
models and are expected to show a rise in the global mean metallicity
with decreasing redshift, if they indeed trace an unbiased sample of
galaxies selected only by $N({\rm H I})$. However, it is unclear whether
or not DLAs actually show such a trend. There has been great debate about
this issue, with most studies advocating no evolution in the global mean
metallicity (Pettini et al. 1997, 1999;  Prochaska \& Wolfe 1999; Vladilo
et al. 2000; Prochaska \& Wolfe 2000; Savaglio 2001; Prochaska 2001;
Prochaska \& Wolfe 2002). In a reexamination of this issue for 57 Zn
measurements in the range $0.4 < z < 3.4$, Kulkarni \& Fall (2002) found
that the slope of the metallicity-redshift relation is $-0.26 \pm 0.10$,
consistent at $\approx$2-3$\, \sigma$ level with model predictions
($-$0.25 to $-$0.61, Pei \& Fall, 1995; Malaney \& Chaboyer, 1996;
Somerville et al. 2001) as well as with no evolution. A similar slope
(-0.25$\pm 0.07$) was also found more recently by Prochaska et al.
(2003b).  

One reason for the ambiguity in the  metallicity-redshift relation is the
large intrinsic scatter in the data.  It is important to determine if the
scatter apparent in the existing samples is cosmic scatter or is caused
by errors of interpretation of the data. It is also important to
observationally investigate whether DLA samples are being affected by
dust obscuration. We address both these issues in this paper. The other
reason for the lack of accurate metallicity-redshift relation is the
small number of measurements available, especially at $z < 1.5$. Most
previous DLA Zn studies have focused on $z > 1.5$ because the Zn II
$\lambda \lambda$ 2026, 2062 lines lie in the ultraviolet (UV) for $z <
0.6$, and likewise for Lyman-$\alpha$ for $z < 1.6$. For $0.6 < z < 1.3$,
the Zn II lines can be accessed with ground-based telescopes, but lie in
blue wavelengths where many spectrographs have lower sensitivity. It is
very important to obtain more data at $z < 1.5$, because this regime
spans  $\sim 70 \%$ of the age of the Universe (for $\Omega_{m}=0.3$,
$\Omega_{\Lambda}=0.7$; $\Omega_m$ and $\Omega_{\Lambda}$ being the
mean comoving matter density and vacuum energy density of the universe,
respectively, in units of the critical energy density of the universe).
The low-$z$ data provide   great leverage on the slope of the
metallicity-redshift relation, and can clarify the relation of DLAs to
present-day galaxies.  A lack of metallicity evolution  would contradict
the high global star formation rate at $1 < z < 1.5$ inferred from the
galaxy surveys such as the Canada France redshift survey, Hubble Deep
Field, as well as the Sloan Digital Sky Survey (SDSS) (e.g. Lilly et al.
1996; Madau et al. 1996; Glazebrook et al. 2003). It would imply that
either DLAs systematically trace only metal-poor dwarf or low surface
brightness galaxies, or that the more metal-rich and dustier DLAs obscure
their background quasars. Clearly, it is necessary to pin down the shape
of the metallicity evolution in a more definitive way.

As a first step toward improving the statistics of the
metallicity-redshift relation at intermediate redshifts, we have recently
started a survey of element abundances for DLAs or strong DLA candidates
(hereafter, CDLAs)  at $z < 1.5$ discovered with the SDSS or the Hubble
Space Telescope (HST). Here we report the results of this survey for a
sample of 11 DLAs with $0.09 < z < 1.5$. Our observations and data
reduction are described in section 2. Section 3 explains the
determination of column densities. Notes on individual objects are
presented in section 4. Section 5 describes our results.
Finally, a discussion of our results and implications for the redshift
evolution of dust content and metallicity are presented in Section 6.

\section{Observations}

Our sample consists of quasar absorbers at $0.09 < z < 1.5$, for which
either a damped Lyman-$\alpha$ line is observed in HST spectra, or a
strong DLA is expected on the basis of Mg II or Fe II lines (Rao \&
Turnshek, 2000) available in the SDSS Early Data Release (EDR) spectra
(Schneider et al. 2002; York et al. 2005a). In particular, we have chosen
systems that have $W_{\rm Mg II 2796}^{\rm rest} > 1.0$ {\AA} and some
other indicator of high $N({\rm H I})$ (Fe I, C I, Mg I, Mn II, Ni II, Ca
II, Cr II, Zn II or Si II $\lambda 1808$). For six of these SDSS systems
$W_{\rm Fe II 2344}^{\rm rest} > 0.79$ {\AA}. In the case of the
$z_{abs}=1.0311$ system in SDSS1727+5302, $W_{\rm Fe II 2344}^{\rm rest} =
0.56$ {\AA}, but this quasar was observed because of the other system at
$z_{abs}=0.9449$ in its spectrum. Objects which met these selection
criteria and which could be observed on our assigned nights constitute
our target list, given in Table 1. The table lists the name of the QSO;
plate number, fiber no. and Mean Julian Date (MJD) of the Sloan Digital
Sky Survey (SDSS) observation used to select the targets; the emission
line redshift of the QSO; the SDSS g magnitude (Fukugita et al. 1996);
the redshift for the DLA targeted in the observation; the grating used
for the Multiple Mirror Telescope (MMT) observations; the FWHM achieved
for the calibration lamp emission lines; the exposure time; and the
signal-to-noise ratio (S/N) achieved per pixel. The overview of the SDSS
project is given by York et al. (2000) and the data products are
described by Stoughton et. al (2002). The camera used for the imaging
observations are described by Gunn et al. (1998). The use of colors
derived from the camera to select QSOs for spectroscopy is described by
Richards et al. (2002). The data are rereduced and republished as new
objects are observed. Data Release 1 (Abazajian et al. 2003) and Data
Release 2 (Strauss et al. 2004) give modifications to the EDR data
formats and products. The spectra used in this paper are on the SDSS
project web site (http://skyserver2.fnal.org). The measurements used for
selecting our targets, as modified by rereductions for inclusion in the
tables below, are from York et al. (2005a).  

Previous data on these DLAs are listed in Table 2. Following a four digit
designation (see Table 1) of the QSO and a two digit version of the
absorber redshift from later tables are given the rest frame equivalent
widths for the strong lines of Fe II in multiplets UV1 and UV3 and Mg II
(UV1), as well as those of some weaker lines we used to select the
systems for observation: Mn II (UV 1); Mg I $\lambda$ 2853 (UV 1); and Ca
II multiplet 1. Where available, the 1 sigma rest equivalent width errors
are listed just below the equivalent widths (even rows 4 thru 18). The
sources are listed in the footnotes.

Our new observations were obtained at the 6.5 m MMT, during October
10-12, 2002 and March 23-25, 2003. Approximately 1.8 nights were lost to
bad weather. The blue channel spectrograph was used with a 832 l/mm
grating in the  second order and a 1200 l/mm grating in the first order.
A CuSO$_4$ order sorting filter was used to block first order red light
when using the 832 l/mm grating in second order. The total wavelength
range covered for 832 l/mm grating was from 3280 {\AA} to 4420 {\AA} and
that for 1200 l/mm grating was 3920 {\AA} to 5420 {\AA}. Each grating was
used with somewhat different central wavelength settings for different DLAs,
depending on its redshift. The dispersions were 0.36 {\AA} per pixel and
0.50 {\AA} per pixel respectively, for the 832 and 1200 l/mm gratings. A
1.0$\arcsec$ wide slit was used. In order to achieve the required S/N
while avoiding cosmic rays, we took several exposures of each object of
durations 1800 s or 2700 s depending on their magnitudes. Each exposure
was preceded and followed by He+Ne+Ar lamp exposures as well as quartz
lamp flat field exposures to get accurate wavelength calibration and flat
fielding. Biases, additional quartz flats, and spectra of bright stars
were taken in the beginning and end of each night.  

The data were reduced using the standard spectrographic reduction
packages of IRAF\footnote{IRAF is distributed by the National Optical
Astronomy Observatories, which are operated by the Association of
Universities for Research in Astronomy, Inc., under cooperative agreement
with the National Science Foundation.}. Instrumental resolution was
obtained from Gaussian fits to the lamp lines near the wavelengths of all
lines to be analyzed in the object spectra. It varies from 1.04 {\AA} to
1.54 {\AA} over most of the wavelength range. The continuum-fitted
spectra with line identifications for the confirmed DLAs and CDLAs are
given in Figs. 1-8. All lines more significant than 3 $\sigma$ in the
DLAs were identified and measured for equivalent widths using the splot
task in IRAF. The absorption systems found are given letter designations
A, B,...,  which are listed beside the redshifts, in increasing order,
across the top of the plots. The figures, in some cases, also include
tick marks at the expected positions of lines discussed later on in the
text, which have not been detected. 

Table 3 lists the atomic data (species, vacuum laboratory wavelength,
oscillator strength and reference for the data source) used for the
identification of lines and for the reductions. 

Table 4 lists the rest-frame equivalent widths, with 1 $\sigma$ errors,
of the various heavy element lines measured for each system. For lines
which are  blended with stronger lines of other systems, we give the
equivalent width of the blend as an upper limit. For weaker (below the 3
$\sigma$ detection limit) lines of those species for which the stronger
lines have been observed, we give the 3 $\sigma$ upper limit on the
equivalent width, obtained assuming the line to be three pixel wide. The
table lists the QSO by the first four digits of its designation in Table
1, and the letter designation of the system; the absorption line
redshift, determined from the average redshifts of the unblended lines of
the system detected at the 4 $\sigma$ level; the species; the rest
wavelength; and the difference between the redshift determined for that
line and the redshift in column 2. This $\Delta z$ is given in units of
10$^{-5}$, so, for instance an entry of 5 at z of 1 corresponds to 7.5 km
s$^{-1}$ or 0.1 {\AA} at 4000 {\AA}, about 1/4 of a pixel width (832 l/mm
grating, 2$^{nd}$ order) or 1/5 of a pixel width (1200 l/mm grating
1$^{st}$ order).

\section{Determination of column densities}

The reasonably high values of S/N for our data made it possible to
perform profile fitting analysis using the package FITS6P
(Welty et al., 1999, 2001). 
 
The lines of Zn II and Cr II are often strong, stronger than those
detected in some other studies. For instance, our average rest frame
equivalent widths of the blend of Zn II and Mg I lines at 2026 {\AA} and
of Cr II line at 2056 {\AA} are 182 m{\AA} and 152 m{\AA} respectively,
while those of Pettini et al. (1994) are 92  m{\AA} for both lines. This
may be partly due to the fact that in order to ensure the DLA nature of
the absorbers, we have chosen systems from the SDSS spectra that have
strong Mg II and Fe II lines and some other indicator of high N$({\rm H
I})$ (C I, Mg I, Si II $\lambda 1808$, Mn II, Ni II, Ca II, Cr II, or Zn
II). We thus could have chosen DLAs at the high column density end (which
after all dominate  the N$_{\rm HI}$-weighted global mean metallicity).
However, in principle, it is also possible that, due to chemical
evolution, the abundances at lower redshift are higher than the
abundances at the redshifts ($>$ 2) observed, e.g., by Pettini et al.
(1994), which may be responsible for the higher line strengths measured
by us. 

At the resolution of 75-95 km s$^{-1}$, it is not possible for us to
resolve the Mg I $\lambda$ 2026 line from the Zn II $\lambda$ 2026 line,
and the Cr II $\lambda 2062$ line from the Zn II $\lambda$ 2062 line.
Therefore, in  determining the column densities of Zn II and Cr II, we
have used information about other lines of Mg I, wherever necessary, and
of Cr II. For Mg I, we have determined the column density from the
equivalent width of the Mg I $\lambda$ 2853 line observed by the SDSS or
by other observers as listed in Table 2, using the $b$ value obtained
from the profile fit as described below. It is possible that the Mg I
column density as obtained from the $\lambda$ 2853 line may be affected
by saturation effects, however, no other line of Mg I is detected. We
have tried to estimate the effect of the uncertainty in the Mg I column
density on the column densities of Zn II as described below. For Cr II
$\lambda$ 2062, we have used the Cr II $\lambda$$\lambda$ 2056, 2066
lines. 

The general method followed for the analysis of the Zn II, Cr II, Mg I
lines between 2025 {\AA} and 2070 {\AA} was as follows. We assumed
the central velocity, $v$, and the effective velocity dispersion, $b$,
of these lines to be same. 

1. The lines of Cr II $\lambda \lambda$ 2056, 2066 were fitted first.

2. The above two lines plus Zn II $\lambda$ 2062 and Cr II $\lambda$ 2062
were fitted simultaneously, using as starting values the results of step
1. The common central velocity, the common effective $b$ values and the
two column densities were all varied to get the best fit.

3. Using the results for the four lines of step 2, the lines were refit,
including Zn II $\lambda$ 2026 and Mg I $\lambda$ 2026. In some cases,
where we could derive the column density of Mg I, we compared it with the
value derived from the SDSS equivalent width of $\lambda$ 2853, to check
for consistency. In others, the data for $\lambda$ 2853 along with the
$b$ value obtained after step 1 and 2, were used to obtain values or
limits that were used for the Mg I $\lambda$ line and the other five
lines were allowed to vary.    

4. Sometimes, when it was not possible to fit the Cr II $\lambda$ 2056
line because of noise or a blend, and so it was not possible to determine
the $b$ value, we took the  $b$  value for the  Zn II, Cr II and Mg I
lines to be the same as that for Si II or for Fe II. In principle the
effective $b$ value for the Fe II or Si II lines could be larger than
that for the Zn II and Cr II lines, if the latter lines  originate in
only a few of the several components contributing to the observed Fe II
and Si II lines. However, we note that these $b$ values are comparable to
or lower than the $b$ values obtained for systems for which we could fit
the Zn II and Cr II lines directly. Furthermore, we have also used the
weak lines of Fe II (which can be comparable in strength to the Zn II and
Cr II lines) whenever available. In any case, we have also
estimated the variation in the Zn II and Cr II column densities by
varying the $b$ values as explained below.

In cases where all the parameters were allowed to vary, the 1 $\sigma$
uncertainties in the parameters were computed simultaneously. In cases
where the effective $b$ values were assumed to be the same as those of Fe
II or Si II lines, we determined the uncertainties in column densities,
due to the uncertainty in $b$, by varying the effective $b$ value by $\pm
1\;\sigma$ as obtained for the Fe II or Si II lines used. We also
investigated the effect of relaxing the constraints of same $b$ value
and same central velocity for lines of different ions i.e. Zn II, Cr II
and Mg I and also, the effect of varying the Mg I column density
(whenever it was taken from the SDSS data) to within its 1 $\sigma$
range. The reported errors in column density are inclusive of these
effects, for all DLAs.   

Table 5 lists the column densities and effective $b$ values with 1
$\sigma$ errors, derived for the various heavy elements from our MMT
data. The QSOs and redshifts are indicated by the same codes used in
Table 2 and 4. We note that the column densities and the equivalent
widths, obtained from them, for all individual lines of Zn and Cr are
found to lie on the linear portion of the curve of growth. The column
densities obtained are, therefore, only marginally dependent on the $b$
values and are not affected by line saturation. We note that this could
partly arise from the large values of $b$ which are the result of the
resolution used for our observations and are indicative of multiple
components. Although it is certainly desirable to obtain higher
resolution observations in the future to resolve these components, the
column densities obtained here should be close to the actual values
(Jenkins 1986) unless there are hidden components with significant column
density but extremely small $b$ values. 

Values of H I column densities, whenever available, and the column
densities for some other ions, which were obtained from previously
measured equivalent widths given in Table 2, are also included in Table
5. 

\section{Notes on Individual Objects}

{\bf Q0738+313} [z$_{em}$=0.630; z$_{abs}$=0.0912, system A;
z$_{abs}$=0.2210, system B] The spectrum in Fig. 1 shows Ti II and Ca II
in system A, the Mg II doublet and a marginal detection of Mg I in system
B. We also place limits on Be II $\lambda$ 3131. The ionization potential
of Ti II is close to that of H I, so N$_{\rm TiII}$/N$_{\rm HI}$ gives
directly the abundance of titanium. For system A, assuming $b$ for Ti to
be the same as that for Ca, we get log(N$_{\rm TiII}$/N$_{\rm HI}$)=-8.7,
giving [Ti/H]$<-$1.6. In this paper, we use the notation [X/H] to mean
the absolute logarithmic abundance of X with respect to hydrogen,
including all forms of H and all forms of X, relative to the solar
abundance. All the abundances given in this paper are upper limits
because we do not take into account the possible presence of H$_2$ and H
II, an assumption also made in all previous DLA abundance studies. Note
that in high redshift DLAs H$_2$/H I ratio has been observed to be small
(e.g. Ledoux, Petitjean \& Srianand 2003 and references therein),
although such high H I columns are almost always accompanied by
comparable H$_2$ in the Milky Way. There is a need to empirically
determine the H$_2$/H I ratio in low and intermediate DLAs, in order to
study more precisely the redshift-abundance relation.

Interstellar Ca II lines arising in the Milky Way are seen in most
spectra and are labeled ``MW'' in the plots.

For Ca II, the ionization potential is much lower than for H I, so
ionization corrections are needed and the interpretation is more
ambiguous than for Ti. The fact that the Ca II lines in the $z=0.0912$
system are a little weaker than the Milky Way (MW) Ca II lines is
consistent with the Ti depletion. The Routly Spitzer effect (Routly \&
Spitzer, 1952) normally exists in such MW sight lines, so the Ca II is
not as depleted as in cold clouds. The calcium in system A may be as
depleted as the titanium. 

In the system B [Ti/H]$<-$1.7. The Mg II lines in this system are
remarkably weak and possibly indicate extreme saturation in individual,
unresolved components. We thus get a lower limit on the Mg abundance as
[Mg/H]$ > -$3.13. 

{\bf Q0827+243} [z$_{em}$=0.941; z$_{abs}$=0.259, system A;
z$_{abs}$=0.5249, system B] System A shows only a weak Mg II system,
similar in strength to that in the $z=0.2209$ system in Q0738+313, but
with strong evidence of saturation.

System B, of primary interest for this paper, contains strong lines
$\lambda\lambda$ 2344, 2374, 2382, 2586, 2600 and weak lines
$\lambda\lambda$ 2249, 2260 of Fe II; and lines $\lambda\lambda$ 2576,
2594 and 2606 of Mn II. A good limit is available for Fe I $\lambda$
2484. Multiple components are clearly indicated in the Fe II lines, with
two dominant components separated by 50 km/sec. Column densities for both
components are given in the table 5. Mn II lines also require two
components with  $b$ values and central velocities very similar to those
of Fe II lines. The 1 $\sigma$ errors in the $b$ values for Mn II are,
however, very large and so we take the $b$ values to be same as that of
the Fe II components. Single component fits require very large value of
$b$ (122 km s$^{-1}$) but give the same column density within the error
bars.
 
We note that the limit on the ratio N$_{\rm Fe I}$/N$_{\rm Fe II}$ ($<
10^{-3}$) is comparable to the limits in our Galaxy in dense clouds: the
radiation field in which the system B is immersed must be as effective in
creating this low ratio as it is in the dense clouds in the disk of Milky
Way.

{\bf Q0933+733} [z$_{em}$=2.528; z$_{\rm abs}$=1.479, system A; z$_{\rm
abs}$=1.4973, system B; z$_{\rm abs}$=1.8283, system C; z$_{\rm
abs}$=1.8573, system D; z$_{\rm abs}$=2.113, system E; z$_{\rm
abs}$=2.333, system F; z$_{\rm abs}$=2.4500, system G; z$_{\rm
abs}$=2.5380, system H] The primary system of interest for this paper is
system A, which includes three, detectable, unblended lines of Ni II, the
weakest permitted line of Si II, the doublet of Ti II, the doublet of Al
III, the Zn II doublet and the Cr II triplet. From SDSS there is an upper
limit of 0.3 {\AA} on the rest equivalent width of the Mg I $\lambda$
2853 (Table 2). We obtain an upper limit of 13.6 m{\AA} on the equivalent
width of Mg I $\lambda$ 1827. The Si II $\lambda$ 1808 yields a $b$
value of 28.6 km s$^{-1}$. The $b$  value obtained for Ni II $\lambda \,
\lambda$ 1741, 1751 is also consistent with this $b$ value within 1
$\sigma$ error. A good fit can be obtained for the Zn II, Cr II, Mg I and
Ti II lines using this $b$  value. The column density of Mg I, which is
treated as a free parameter, is much smaller than the upper limit of $1.9
\times 10^{13}$ cm$^{-2}$ obtained from the absence of 1827 line in our
spectra and is also lower than the upper limit obtained from the
equivalent width limit in Table 2. The Mg I column density is not well
determined by the profile fit and the 1 $\sigma$ errors are much larger
than the fitted value. We have, thus, only given the 3 $\sigma$ upper
limit. 

As for the other systems, systems C, D, G and H are seen only in C IV and
not, for instance, in Si II ($\lambda$ 1526) or Al II ($\lambda$ 1670),
the strongest lines of first ions in the observed spectral region. The
absence of Al II in system C suggests that the possible Si II line of
system C, blended with Ni II $\lambda$ 1741 of system A is of no
consequence and that the line is fully Ni II of system A. System E is a
multi-component system seen in C IV but also detected in Si IV and
definitely not seen in Al II. System B is seen only in the Al III
doublet, only 2200 km s$^{-1}$ from the same doublet in system A but
showing no trace of the many other lines of system A, even though Al III
is much stronger in system B than in system A.

System F has very strong C II, C IV, and Si IV; detectable Si II
$\lambda\lambda$ 1526, 1304; excellent limits on Fe II $\lambda$ 1608,
N$_{\rm SiII}$/N$_{\rm Fe II}>$ 21.1; but no detectable O I $\lambda$
1302: O I could be apparently missing because it is blended with Si IV in
system E, but this is unlikely since the Si IV profile mimics well the
multi-component C IV profile of the same system, showing no evidence of
other structure that could be attributed to O I. System F is thus a
purely ionized region, having high ions, of very low density, judging
from the absence of CII$^*$ (observed wavelength 4451.9 {\AA}). The
absence of O I and C I in system F is in sharp contrast to the system B,
in SDSSJ2340-0053 discussed below, in which O I is very strong in the
presence of a similar set of ions.

There are two unidentified lines at 4515 and 5436 {\AA} and possibly a
third one at 5461 {\AA} (blended with C IV in system H) which could most
likely be the C IV 1548 line of additional systems at $z=1.917$, 2.511
and 2.528 in which the C IV 1550 line is too weak to detect.

The remaining objects are quasars from the SDSS. Only SDSSJ1028-0100 has
been observed before for absorption lines (Petitjean et al. 1998,
hereafter P98). The objects are included in the DR1 list of QSOs
(Schneider et al. 2002) The original SDSS spectra from which these
objects were selected are on the public website:
http://skyserver.sdss.org/dr1/en/. The catalog of lines found in SDSS is
given by York et al. (2005a) and is partially included in Table 2. All
the systems found below can either be identified in the SDSS spectra or
are too weak to be detected in the SDSS QSO survey spectra. To use the
website, the coordinates of the objects can be used, or one can use the
plate and fiber numbers in Table 1. 

{\bf SDSSJ 1028$-$0100} [z$_{em}$ = 1.542; z$_{\rm abs}$ = 0.324, system
A; z$_{\rm abs}$ = 0.6321, system B; z$_{\rm abs}$ = 0.7088, system C;
$z=1.265$, system D; z$_{\rm abs}$ = 1.484, system E] This QSO was
observed by P98 (Q1026-0045B, in their paper) with HST, along with the
nearby object Q1026-0045A (z=1.438)(SDSSJ 102835-010043.7). 

Our data show system A to be a multicomponent Mg II system, with two Fe
II lines in the observed range. The Fe II lines are weak and we infer
N$_{\rm FeII}$/N$_{\rm MgII}$ = 2.02. Mg I $\lambda$ 2853 is present.

Systems B and C are the main systems of interest in this paper. The two
systems are, interestingly, separated in $z$ only by a little more than
the two damped systems in SDSSJ1727 (Turnshek et al., 2004). Additional
data are available from P98 as given in Table 2.
 
For system B, the Zn II and Cr II lines are in noisy part of the spectrum
and we used the $b$ value of Fe II lines for these lines. The column
density of Mg I was obtained for this value of $b$  from the $\lambda$
2853 line observed by the SDSS. The errors on the estimated column
densities of Zn and Cr are large because of the poor quality of the
spectrum in the region of interest. A good limit is obtained for Fe I
$\lambda$ 2484.

For system C Zn II $\lambda$ 2026 and Mg I $\lambda$ 2026 lines are
blended with Si IV $\lambda$ 1393 of system E. The $\lambda$  2056 line
of Cr II is blended with $\lambda$ 1550 line of C IV of system D. The $b$
value for the Zn II and Cr II lines was assumed to be the same as that
for Fe II. The column density of Mg I was obtained from the SDSS data for
this value of $b$. Parameters for C IV lines of system D and Si IV lines
of system E were obtained by fitting the unblended lines of these systems
first. All the lines were then fitted together. Even though we were able
to get the column density of Cr II reasonably well the column density of
Zn II is poorly determined. 

The HST spectrum for this QSO has become available recently (HST GO project
No. 9382, P.I. Rao, S.). For both CDLAs the column densities are found to
be close to $10^{20}$ cm$^{-2}$ and the systems thus fall into the
category of sub-DLAs. We have included our estimates of the H I column
densities obtained by fitting the line profile to the HST data in Table
5. The 1 $\sigma$ errors in H I column densities were estimated to be
$\approx$ 30 $\%$ taking into account the uncertainties in fitting the
continuum. Because of the sub-DLA nature of these systems, we do not
include them in our calculations presented in sections 5.2 and 5.4.

The pair of QSOs (Q1026-0045A and Q1026-0045B) contain several common
systems (nearly identical redshifts), though they are separated by $\sim
300$ kpc. One is a pair of Lyman $\alpha$/Lyman $\beta$ systems at $z =
1.4439$ in object B and 1.4420 in object A, in which no lines of heavy
elements are found. We have very good limits on Si IV, C IV, Al II, Si II
and Al III: none of these features are seen to our limit of detection. 

Another common pair is the system at $z = 1.265$ (our system D). Lyman
$\alpha$ and Lyman $\beta$ were seen in both objects. Si IV $\lambda$
1393 was detected in object A. We confirm the presence of strong C IV in
the object B system, but can set stringent upper limits to the strengths
of line of Al II, Si II and Al III. Evidently, this is a high ionization
system, that must be larger than 300 kpc in transverse dimension as noted
by P98. In object A, there is a separate absorber at $z = 1.2969$ which
has O I, numerous first and second ions and Si IV. No lines are detected
at this redshift in object B (our spectrum), so this is not a common
pair.

System E was found (P98), in object B (not seen in object A), to have no
first ions. It shows C IV and Si IV in the MMT spectrum, and N III, C
III, Si III and possibly O VI in the spectrum from HST (P98). Again, no
low ions (Al II, Si II) are present to strong limits in the MMT spectrum,
and Al III is not present. This is a low column density, highly ionized
system.

We find one system, A, at $z=0.324$ in Fe II $\lambda$ $\lambda$ 2586,
2600. The other three, strong Fe II lines are clearly present in the P98
spectrum. One of these, and a weak C IV system at $z=1.01$ led P98 to
suggest, tentatively an Mg II doublet at $z=0.11$: there is no evidence
that this system is real as it is fully explained by the weak C IV
doublet and our system A. Two lines of Fe II, two of Mg II and one of Mg
I are detected. The lines of Fe II and Mg II are weak and we infer
N$_{\rm Fe II}$/N$_{\rm Mg II}$=2.

In addition to the systems noted above, we can place stringent upper
limits on any possible Al II for eleven Lyman $\alpha$-only systems
between $z=0.96$ and 1.5. 

{\bf SDSSJ 1107+0048} [z$_{em}$ = l.302; z$_{\rm abs}$ = 0.7405, system
A; z$_{\rm abs}$ = 1.0704, system B; z$_{\rm abs}$ = 1.3503, system C;
z$_{\rm abs}$ = 1.3687, system D] The prime system is A. In this system,
two-components are present in Fe II, Zn II, and Cr II lines. For Fe II
$\lambda$  2249 and Cr II $\lambda$  2066, a two-component fit gives the
same column densities as does a single-component fit, with similar rms
values for the fit. As there are too many variables for a two component
fit of the Zn II, Cr II and Mg I lines, we used single components to get
the column densities. The Mg I column density, which was treated as a
free parameter, is consistent with the value from SDSS for the $b$
value obtained here, within 1 $\sigma$ limits. Upper limits on Co II
and Ti II are from the absence of $\lambda$ 2012 and $\lambda$ 1910 lines
respectively. The HST spectrum for this QSO has become available recently
(HST GO project No. 9382, P.I. Rao, S.). We have included our estimate of
the H I column density obtained from the HST data in Table 5. The 1
$\sigma$ errors in H I column densities were estimated to be $\approx$ 30
$\%$ taking into account the uncertainities in fitting the continuum. 

System B contains only Al III and Al II, with comparable strengths.
Systems C and D are weak C IV systems with no other lines detected.

{\bf SDSSJ 1323$-$0021} [z$_{em}$ = 1.390; z$_{\rm abs}$ = 0.7160, system
A; z$_{\rm abs}$ = 1.3574, system B; z$_{\rm abs}$ = 1.3742, system C;
z$_{\rm abs}$ = 1.3868, system D; z$_{\rm abs}$ = 1.3891, system E]

System A, is the main system for our purpose, for which we have lines of
Zn II, Cr II, and Fe II and limits on Co II. The $\lambda$ 2056 line of
Cr II is noisy and so can not be fitted to get the $b$  value and the
column density of Cr II. Therefore, the value of $b$ is assumed to be
the same as that of Fe II. The column density of Mg I is obtained from
the SDSS data for this $b$. 

Systems B, C, D and E are all C IV only systems, all within 4000 km/s of
each other. For systems C, D and E, Si IV is in the observed range, but
is below the detection limits. 

The HST spectrum for this QSO has become available recently (HST GO project
No. 9382, P.I. Rao, S.). We have determined the H I column density by
fitting the line profile to the HST data and find it to be smaller than
2$\times 10^{20}$ cm$^{-2}$. The system thus falls into the category of
sub-DLAs. We have included our estimate of the H I column density in
Table 5. The 1 $\sigma$ errors in H I column densities were estimated
taking into account the uncertainties in fitting the continuum. Because
of the sub-DLA nature of this system, we do not include it in our
calculations presented in sections 5.2 and 5.4.

{\bf SDSSJ 1727+5302} [ z$_{em}$ = 1.444; z$_{\rm abs}$ = 0.9449, system
A; z$_{\rm abs}$ = 1.0311, system B; z$_{\rm abs}$ = 1.2565, system C;
z$_{\rm abs}$=1.3064, system D] Systems A and B are the prime systems for
this paper. These systems reveal weak lines of Ni II, Al III, Si II, Cr
II and Zn II. Ti II is detected in  system A. A strong line is present at
the position $\lambda$ 1941 of Co II. However, the line $\lambda$ 2012 of
Co II, having comparable $f$ value, is absent and so this line must be a
blend. Limits are placed on Ti II and Co II column densities in system B.
For system A, the column density of Mg I that we get by fitting our data
is consistent within 1 $\sigma$ with the value obtained from SDSS data,
within error bars. Other singly ionized species also give $b$ values
similar to Zn II, Cr II, Mg I lines. For system B a good fit is obtained
for $b$ value close to those of the other singly ionized lines. Mg I is
taken from the SDSS value.  

For systems A and B we adopt the H I column densities given by Turnshek
et al. (2004) but allow for 30\% error in these values in light of the
confusing region (Lyman beta emission line) in which the damped systems
appear and the several choices of continuum that are possible for the
region.

Systems C and D are weak, C IV-only systems. 

{\bf SDSSJ 2340$-$0053} [z$_{em}$ = 2.085; z$_{\rm abs}$ = 1.3606, system
A; z$_{\rm abs}$ = 2.0547, system B; z$_{\rm abs}$ = 2.062, system C]
System A is the prime system for this study. In this system, we detect Al
II, Ni II, Si II ($\lambda$ 1808), Zn II, Cr II and Fe II. The Mg I
column density obtained here is consistent with the SDSS value within the
1 $\sigma$ error. A limit is obtained for Al I $\lambda$ 2263.

System B has a broad range of ionization states represented, from C I to
C IV. System C has only C IV and is separated by only 200 km/sec from B,
so the two systems are probably associated with each other and with the
quasar. Note that the strong O I line indicates that three regions of
different ionizations are involved in system B/C: an H I region (dominant
in H I and O I with traces of the C I and presumably with Si II, C II and
Al II); and an H II region, possibly with a wide range of ionization (C II,
C IV, Si II, Si IV, Al II, Al III); and a highly ionized region, system
C, with C IV only (and no Si IV in particular). The fine structure
excited line, C II* $\lambda$ 1335 may be present, but it is ambiguous
whether it is from the H I region, the H II region, or both. The system is
similar to what one sees looking from afar at a QSO through Loop I
(Burkes et al. 1991) and the Milky Way. The C I line at 5060 {\AA} shows
evidence of the C I* and C I** components, implying significant densities
in the H I region. The H I region may be a good candidate for detecting
molecular hydrogen: the ratio N$_{\rm SiII}$/N$_{\rm FeII}$ is  low, as
in  cold cloud gas in our Galaxy, indicating the
presence of grains, and the neutral gas is cold. These are the main
conditions for forming H$_2$ (Tumlinson et al. 2002). On the other hand,
the Cl I 1347 line, formed by charge exchange between Cl$^+$ and H$_2$ and
normally strong in H$_2$-containing clouds in our Galaxy (Jura \& York
1978), is not present. Either the overall abundance of Cl is low, or the
amount of H$_2$ is low, or both.




\section{Results}

\subsection{Relative Abundances}

Table 6 lists the abundances of Cr, Fe, Ni, Al, Si, Ti, Mn, and Co relative
to Zn, and Si relative to Fe, for the systems in our sample. We have used
solar system abundances from Lodders (2003) as a reference. The table
also lists the averages of these relative abundances as well as the
average values obtained for the compilation of Prochaska et al. (2003a)
for DLAs at $z>2.5$. The values for Galactic warm clouds, cold clouds and
halo clouds, taken from Welty et al. (1997,1999a,1999b,2001) and York et
al. (2005b) are also included for comparison. 

The relative abundances of various elements are determined both by the
nucleosynthetic processes and by the differential dust depletion. If the
abundance pattern resembles that of Galactic halo stars, the $\alpha$
elements such as Mg, Si, S, O, Ar may be nucleosynthetically enriched
compared to the iron group elements such as Cr, Mn, Fe, Co, Ni and Zn
(because the former are produced by Type II supernovae). The odd-Z
elements such as Al and Mn may be deficient compared to even-Z elements
such as Si and Fe, respectively. Furthermore, the refractory elements
such as Si, Ti, Cr, Mn, Fe, Co, and Ni are expected to be depleted if the
DLAs contain dust. We use Zn as the reference element because, as noted
before, it is almost undepleted in warm Galactic interstellar clouds and
shows relatively small depletion even in cold Galactic interstellar
clouds. [Cr/Zn] is a measure of the amount of dust, and we discuss it in
detail in the next subsection. 

For both sub-DLAs toward SDSS1028-0100, the errors in Zn II column
densities are large. We, therefore, use the 3 $\sigma$ upper limits on
the Zn II column density and give lower limits to the abundances of
various elements w.r.t. Zn for these systems. The average values of
relative abundances as measured by us are very similar to the values
obtained by Prochaska et al (2003a) at redshifts $>$ 2.5, suggesting that
the nature and amount of dust in DLAs do not vary much with redshift. 

For [Si/Zn] Prochaska et al. (2003a) have reported only one measurement
of 0.08. Even though this is different from the average value obtained by
us, it is well within the spread of our values. Overall our values,
together with those compiled by Ledoux, Bergeron \& Petitjean (2002a),
Prochaska et al. (2003a) and Dessauges-Zavadsky, Prochaska \& D'Odorico
(2002) (a total of 18 in all), show [Si/Zn] to be in the range of -0.8 to
+0.2. If the depletion pattern in DLAs is similar to that in the warm
Galactic clouds, these values indicate super-solar total abundance of Si
by a factor of 2 or larger in only 3 of these systems. 

Welty et al. (2001) note, however, that [Si/Zn] is solar in the gas in
the Small Magellanic Cloud (SMC) toward the star Sk 155, with a
so-called SMC extinction curve. Other stars in the SMC show the same
effect (Welty et al. in preparation), even though [Fe/Zn] is subsolar in
the same sight lines. Thus, one must allow for the possibility that the
grain composition differs from place to place and that Si is just not
depleted much in some cases. In that case, the abundance of Si may not be
as high as noted above.

Ti has been observed in very few DLAs so far. Dessauges-Zavadsky
et al. (2002) and Prochaska et al. (2003a) have reported one value each,
0.29 and -0.94, respectively, for [Ti/Zn], while the average of 5 values
compiled by Ledoux et al. (2002a) is -0.26. We measure [Ti/Zn] in two
systems. Altogether, only 3 out of 9 of these values indicate super-solar
abundance by a factor of 2 or larger, assuming the depletion to be
similar to that in warm Galactic clouds. This suggests that $\alpha$
enrichment by type II supernovae may not be important in a large fraction
of the DLAs.

We have measured the column densities of Al II and Al III in two systems
which have Zn measurements. For one of these we also have an upper limit
on Al I which is much smaller than the column densities of the other Al
ions. Thus we have [Al/Zn] = -1.2 and -0.6 for $z=1.03$ DLA in
SDSSJ1727+5302 and $z=1.36$ DLA in SDSSJ2340-0053 respectively. This
suggests dust depletion in these DLAs, irrespective of whether the
intrinsic nucleosynthetic pattern is similar to the solar pattern or that
observed in halo stars (e.g. Lauroesch et al. 1996; Kulkarni, Fall \&
Truran 1997). However, we note that Al II lines are strong and may
possibly be saturated, in which case these values should, strictly, be
treated as lower limits.

We have measured [Mn/Zn] in one system to be -0.6. The average of 9
values compiled by Ledoux et al. (2002a) is -0.66 and that of the three
values reported by Dessauges-Zavadsky et al. (2002) is -0.77. All these
values suggest moderate to high depletion for the solar abundance pattern
as well as the halo-star abundance pattern.

The values of [Ni/Zn] obtained by us indicate substantial depletion for
the solar abundance pattern as well for the halo-star abundance pattern.

\subsection{Redshift Evolution of Dust Content}

The [Cr/Zn] values for our sample are similar to the [Cr/Zn] values found
at high redshifts. To make a quantitative comparison, we combined our
data with data from the literature. We examined the data of Boisse et al.
(1998); Centurion et al. (2003), Dessauges-Zavadsky et al. (2003, 2004);
de La Varga et al. (2000); Ellison \& Lopez (2001); Ge, Bechtold, \&
Kulkarni (2001); Ledoux, Srianand, \& Petitjean (2002b); Ledoux,
Petitjean, \& Srianand (2003); Lu et al. (1995, 1996); Lopez et al.
(1999, 2002); Lopez \& Ellison (2003); Meyer \& York (1992); Meyer,
Lanzetta, \& Wolfe (1995); Molaro et al. (2000); P\'eroux et al. (2002);
Petitjean, Srianand, \& Ledoux (2000, 2002); Peroux et al. (2002);
Pettini et al. (1994, 1997, 1999, 2000); Prochaska \& Wolfe (1996, 1997,
1998, 1999, 2000); Prochaska, Gawiser \& Wolfe (2001a), and Prochaska et
al., (2001b, 2003a, 2003b, 2003c) . To be conservative, we excluded
sub-DLAs (10$^{19}$ cm$^{-2}$ N$_{\rm HI} < 2 \times 10^{20}$ cm$^{-2}$)
and the systems separated by $<$ 3000 km s$^{-1}$ from the QSO redshift.
We, also, excluded cases where only upper or lower limits are available
for Cr and/or Zn, and scaled the measurements to the same set of
oscillator strengths for the Zn II and Cr II lines which are given in
Table 3. In Fig. 9, we plot [Cr/Zn] vs. redshift for this combined
sample of 41 DLAs. Our MMT data have provided 4 new [Cr/Zn] measurements
for DLAs at $z < 1.5$ and nearly doubled the existing sample of [Cr/Zn]
measurements at these redshifts. In addition our MMT data have provided
[Cr/Zn] ratio for two CDLAs and two sub-DLAs which are also shown in Fig.
9.

A linear regression fit to the unbinned [Cr/Zn] vs. redshift data gives a
slope of $0.07 \pm 0.01$, and an intercept of $-0.48 \pm 0.02$
implying a slow decrease in [Cr/Zn] with decreasing redshift. The
linear correlation coefficient between [Cr/Zn] and $z$ is 0.19. 

To examine the extent to which the above conclusion is affected by the
scatter among the individual [Cr/Zn] values, we also binned the data into
six redshift bins with roughly equal number of systems per redshift bin.
The unweighted mean $\langle$~[Cr/Zn]~$\rangle$ in each bin vs. the
median redshift of the bin is plotted in Fig. 10a. The vertical error bar
for each bin denotes the 1 $\sigma$ uncertainty in the unweighted mean,
and includes the sampling uncertainties as well as the measurement
uncertainties calculated by propagating the errors in individual Cr and
Zn measurements. Linear regression for the mean [Cr/Zn] vs. redshift
gives a slope of $-0.03 \pm 0.06$,  an intercept of $-0.26 \pm 0.13$,
and a linear correlation coefficient of $-0.18$. 

To examine the evolution of ``global depletion'', we define the $N_{\rm
Zn II}$-weighted mean $[\langle {\rm Cr/Zn} \rangle ] \equiv \rm{log}
(\Sigma N_{Cr II}^{i} / \Sigma N_{Zn II}^{i}) - \rm{log}
(Cr/Zn)_{\odot}$. Note that $\rm \Sigma N_{Cr II}^{i} / \Sigma N_{Zn
II}^{i}$ corresponds to $\Omega_{\rm Cr}/\Omega_{\rm Zn}$, $\Omega_{\rm
Cr}$ and $\Omega_{\rm Zn}$ being the mean comoving density of Zn and Cr,
respectively, in units of the critical energy density of the universe.
Fig. 10b shows this weighted mean $[\langle {\rm Cr/Zn} \rangle ]$ vs.
median redshift in each bin for our combined sample. The vertical error
bar for each bin denotes the 1 $\sigma$ uncertainty in the weighted mean,
and includes the sampling uncertainties as well as the measurement
uncertainties. The procedures for calculating these uncertainties are
similar to those employed in Kulkarni \& Fall (2002).  Linear regression
analysis for this N$_{\rm Zn II}$-weighted mean depletion vs. redshift
gives a slope of $0.05 \pm 0.06$, an intercept of $-0.41 \pm 0.12$, and a
linear correlation coefficient of 0.32. 

The intercept value for the N$_{\rm Zn II}$-weighted mean depletion is
negative at $ \approx  3.3 \, \sigma$ level, suggesting that the mean
global dust depletion of DLAs is significant at all redshifts examined
($0.6 < z < 3.4$). However, the slope values are consistent with zero.
Overall, we conclude that the existing data do not suggest much evolution
of the conventional dust content indicator, [Cr/Zn] in DLAs with
redshift. Recently Vladilo (2004) has found evidence for an increase of
dust-to-gas ratio with increasing metallicity, using [Fe/Zn] in 38 DLAs,
but has found only a weak evidence for the evolution of this ratio with
time.

\subsection {Quasar Reddening and Element Depletions}

If the observed depletions of elements such as Cr or Fe relative to Zn
are caused by dust in the DLAs, the spectra of background quasars may be
expected to be reddened by the intervening dust. Pei, Fall, \& Bechtold
(1991) found the spectra of quasars with DLAs to be systematically
reddened with respect to the spectra of quasars without DLAs. Similar
study with a much larger, homogeneous sample of QSOs from the SDSS DR2 by
Murphy and Liske (2004) has found no evidence for such reddening at the
redshift of about 3. However, the dust effects are likely to be more
important at the lower redshifts sampled by our DLAs.  Although our
sample is small, it consists of strong DLAs or CDLAs. Therefore, it is
interesting to ask whether the quasars in our sample are systematically
more reddened than a typical quasar without a foreground DLA. To address
this, we focus on the SDSS objects in our sample, for which a uniform set
of measurements are available for the quasar spectral and photometric
properties. Below we make a preliminary search for correlations between
the quasar reddening and [Cr/Zn] in the absorbers, using two different
estimates of the reddening based on (a) the shape of the quasar spectrum,
and (b) the quasar $g-i$ color.

\subsubsection{Reddening Estimates based on Quasar Spectra} 

Reichard et al. (2003) have constructed a composite quasar spectrum using
3,814 spectra from the SDSS EDR. They have also fitted the spectra of the
individual EDR quasars to the composite spectrum, assuming that the
continuum follows a power law $f_{\nu} \propto \nu^{\alpha}$, that is
diminished by dust extinction characterized by reddening $E(B-V)$. They
assume the reddening law for the SMC from Pei (1992). Table 7 lists, for
the SDSS quasars in our sample, the best-fitting values of the spectral
index $\alpha_{RP}$ and the reddening $E(B-V)_{RP}$ as found by Reichard
et al. (private communication). Here, R stands for reddening law and P
for power law, both of which were used to determine these parameters.
Also listed in Table 7 are qualitative  comments about the colors of the
quasars based on visual inspection of the spectra. But we note that there
is a degeneracy between the values of $\alpha_{RP}$ and $E(B-V)_{RP}$,
i.e. several combinations of the two fit parameters can provide almost
equally satisfactory fits. As a result, the $\alpha_{RP}$ and
$E(B-V)_{RP}$ values sometimes have unphysically high or low values. For
this reason, reliable values of $\alpha_{RP}$ and $E(B-V)_{RP}$ are not
available for SDSSJ1028-0100. Two of the remaining quasars
(SDSSJ1107+0048 and SDSSJ2340-0053) are not significantly reddened, and
show $E(B-V)_{RP}$ consistent with the SDSS composite for
``flat-spectrum'' quasars (sample 1 of Richards et al. 2003). The
remaining two quasars (SDSSJ1323-0021 and SDSSJ1727+5302) have higher
values of $E(B-V)_{RP}$. 

\subsubsection{Reddening Estimates Based on Quasar Colors}

We also attempted to constrain the reddenings in another way for the SDSS
quasars, since the $E(B-V)_{RP}$ values may not always be the best
physical indicators of the actual reddening. To do this, we used the
observed $g-i$ colors of the quasars and calculated the excess $\Delta
(g-i) \equiv (g-i) - (g-i)_{med}$, where $(g-i)_{med}$ is the median
$g-i$ color for the SDSS quasar composite at the redshift of the quasar,
taken from Table 3 of Richards et al. (2001). We then use the SMC
reddening law, A$_\lambda = 1.39 \lambda^{-1.2} E(B-V)$ (Prevot et al.
1984) to obtain the $E(B-V)_{g-i}$ from the observed $\Delta(g-i)$
values. Taking $\lambda_g$ and $\lambda_i$ to be 4657.98 {\AA} and
7461.01 {\AA} respectively, this gives $E(B-V)_{g-i} = {\Delta(g-i)
(1+z)^{1.2} /1.506}$, z being the redshift of the absorber. To obtain the
$E(B-V)$ for individual DLAs toward SDSSJ1727+5302 we assumed them to be
in proportion to the H I column densities in these systems. The
$E(B-V)_{g-i}$ values are given in Table 7.

\subsubsection{Search for Correlations between Reddening Estimates and
DLA Depletions}

It is not clear whether the reddening in any of our targets (estimated by
either of the above methods) is caused by dust intrinsic to the quasar or
dust in the intervening absorbers. Indeed, the 2200 {\AA} bump has been
observed in a few quasar spectra at the redshifts of the quasar and/or
the intervening absorbers (Wang et al. 2004). Hopkins et al. (2004),
using SDSS data, have shown the reddening to be dominated by SMC type
dust at the QSO redshifts. In any case, the observed $E(B-V)_{RP}$ or
$E(B-V)_{g-i}$ values can be used to constrain the amounts of reddening
that could be caused by the intervening DLAs. From these, assuming the
$E(B-V)$ to be caused by the DLA absorber and assuming it to be equal to
N$_{\rm{HI}} / D$ (ignoring possible presence of H$_2$) one can either
obtain the H I column densities in the absorbers assuming a value of the
constant $D$ for a specific extinction curve, or obtain the value of $D$
if N$_{\rm{HI}}$ is known.  These column densities, together with the
observed Zn II column densities, can then be used to constrain the
metallicities for the DLAs for which N$_{\rm{HI}}$ is not yet known.
Table 8 lists the N$_{\rm{H I}}$ values inferred from HST spectra (where
available), along with estimates of N$_{\rm{H I}}$ from the reddening
$E(B-V)_{g-i}$, assuming D = 4.6$\times 10^{22}$ cm$^{-2}$ mag$^{-1}$ for
the SMC extinction curve. The table also includes the corresponding
values/estimates of the abundances of Zn and Ti. We note, however, that
the H I column densities obtained from the $E(B-V)$ values for
SDSS1107+0048 and SDSS1323-0021 differ considerably from their HST
values. The H I column densities obtained from the $E(B-V)$ can,
therefore, only be taken to be tentative values. We, thus, do not include
the metallicity obtained in this manner for SDSSJ2340-0053 in our
calculations for metallicity evolution, presented in the next subsection.
Using the HST values of H I column densities we get the values for D of
2.0$\times 10^{22}$, 1.0$\times10^{21}$ and 1.1$\times 10^{23}$ for the
DLAs in SDSS1107+0048, SDSS1323-0021 and  SDSS 1727+5302 respectively.  

Fig. 11 plots the two estimates of reddening for the SDSS quasars in our
sample vs. the [Cr/Zn] in the corresponding DLAs. The circles show the
$E(B-V)_{RP}$ values while the squares show the $E(B-V)_{g-i}$ values.
The filled symbols are for quasars   with single DLAs, while the unfilled
symbols are for a quasar (SDSSJ1727+5302) with two DLAs along the line of
sight, treated as described above to estimate the individual
contributions of the two systems. There may be some anti-correlation
between [Cr/Zn] and the reddening, such as may be expected if the
reddening arises primarily in the absorbers, and the absorbers with the
higher depletion of Cr are also the ones with the greater reddening, and
vice versa. The correlation is less clear if the double sight line is
included, but the reddening estimates for such absorbers are less certain
because of the assumption of the same extinction law for both systems in
the sight line. Clearly, it is necessary to obtain element depletions in
many more SDSS sight lines and to employ more robust estimators of the
reddening to determine whether the correlation possibly present in Fig.
11 is truly significant. 

\subsection{Constraints on Metallicities and Implications for Metallicity
Evolution}

Finally, we study the implications of our MMT data for the metallicity
evolution of DLAs. The analysis presented here is based on the
methodology outlined in Kulkarni \& Fall (2002) and uses the data
compiled therein along with the more recent data referred to in section
5.2 as well as the data presented in this paper. We binned the combined
sample in redshift and calculated the global N$_{\rm H I}$-weighted
metallicity in each bin. 

The last two columns of Table 8 give the estimated abundances of Zn and
Ti. In most systems the abundances of Zn and Ti are smaller than a tenth
of the corresponding solar abundances except in the case of
SDSS1323-0021, for which the Zn abundance has a very high value. 

For the sample of 41 DLAs mentioned above (section 5.2), the $N_{\rm H
I}$-weighted mean metallicity in the $0.7 < z < 1.5$ range is $-1.09 \pm
0.20$, and the linear regression slope of the metallicity-redshift
relation is $-0.21 \pm 0.10$ for the redshift range 0.7 $ < z< $ 3.4.
Considering only  Zn measurements, irrespective of whether or not Cr was
measured, a total of 83 DLAs (51 detections, 32  limits) are currently
available from our MMT data and data from the literature. We constructed
two samples for these 83 DLAs: a ``maximum limits'' sample where the Zn
limits are treated as detections, and a ``minimum limits'' sample, where
the Zn limits are treated as zeros. For an individual system these
extreme cases cover the full range of possible values the Zn column
densities can take in the case of the limits. The $N_{\rm H I}$-weighted
mean metallicity in the redshift range $0.4 < z < 1.5$ is $-1.03 \pm
0.14$ for both the ``maximum limits'' sample and the ``minimum limits''
samples. The linear regression slope of the metallicity-redshift relation
for the redshift range $0.4 < z < 3.9$ is $-0.13 \pm 0.08$ for the
``maximum limits'' sample, and $-0.27 \pm 0.10$ for the ``minimum
limits'' sample. The corresponding estimates for the intercept of the
metallicity-redshift relation are $-0.87 \pm 0.19$ for the ``maximum
limits'' sample and $-0.66 \pm 0.22$ for the ``minimum limits'' sample. 
 
Thus our data seem to indicate that there is no strong evolution of the
global metallicity with redshift. We note that we have only one value at
$z < $0.6 in our data set. Measurements at lower redshifts are important
for better determination of the global metallicity evolution. Kulkarni et
al. (2004) have taken steps in this direction by obtaining HST data at $z
< 0.6$.

\section{Discussion} The issues of dust depletion and associated
selection effects, and the evolution of metallicity in DLAs are not yet
fully understood. The large samples of DLAs or CDLAs becoming
available from the SDSS can be used to improve the statistics of both
element abundances and dust reddening. In this paper, we have presented
results for 7 CDLAs discovered in the SDSS and have doubled the
Zn and Cr samples at $z < 1$. Our data suggest weak evolution of1
metallicity and dust content in the absorbers with redshift.

Kulkarni \& Fall (2002), using 57 Zn measurements in the redshift range
$0.4 < z < 3.4$, found the slope of the $N_{\rm H I}$-weighted mean
metallicity-redshift relation to be $-0.26 \pm 0.10$. A very similar
value of the slope was also obtained by Prochaska et al. (2003b) using
metallicity estimates for 121 DLAs with $0.5 < z < 4.7$. Ten of these are
based on Zn, six on Fe (corrected by adding an assumed level of dust
depletion), one on the X-ray absorption, and the remaining on Si, S, or O.
The estimates of the metallicity-redshift relation presented
in our study are also consistent with both of these results. 

We note that we have used only Zn as a tracer of  metallicity for several
reasons: (a) This avoids complications (present in the case of Si, S, or
O) of significant nucleosynthetic differences from the common metallicity
indicator Fe. (b) Zn allows direct metallicity estimates nearly free from
dust depletion corrections which can be significant for Fe. (c) The use
of one element also makes the study of metallicity evolution more
uniform. Nevertheless, it is very interesting that results based on Zn
alone agree with those based on other elements.
 
We also note that out of the 121 DLAs in the sample of  Prochaska et al.
(2003b), only 9 have $z < 1.5$. Furthermore, the metallicity estimate for
the lowest redshift system in that sample ($z = 0.52$) is based on X-ray
absorption measurement of Junkkarinen et al. (2004), not on Zn, and
there could be systematic differences between metallicity estimates based
on Zn and those based on X-ray absorption. Thus, our MMT measurements at
$0.6 < z < 1.5$ are a significant addition to the existing abundance data
at intermediate redshifts. 

In a recent study, Nestor et al. (2003) have averaged SDSS spectra at the
positions of the Zn II and Cr II lines, all deredshifted to the rest
frame of the absorber, as found from the Mg II lines. They found [Cr/Zn]=
-0.44$\pm{0.13}$ at redshifts averaged from 0.9 to 2.2. These values are
similar to the values in the literature, summarized above, and also with
our values obtained from the analysis of individual spectra of SDSS
objects. However, the variation of the ratio [Cr/Zn] from system to
system is large, as we confirm here for $z<$1.5, and as has been observed
for $z>1.5$. It is important to determine if this variation is caused by
a variation in the dust-to-gas ratio and if that correlates with
variation in the extinction. 

Overall, our observations have demonstrated the potential of the SDSS,
combined with higher resolution follow-up spectra, to study the chemical
evolution of the DLAs. Further studies of a large number of SDSS DLAs at
$z < 1.5$ will help to determine the low-$z$  end of the cosmic
metallicity-redshift relation to a much higher accuracy than before. Such
studies will help to understand whether or not DLA observations agree
with predictions of cosmic chemical evolution models, and whether dust
extinction effects are significant. 







\acknowledgments

We are grateful to the staff at the MMT for technical assistance. We also
thank T. Reichard, P. Hall, and G. Richards for discussions about
reddening for the SDSS quasars. PK acknowledges travel support from the
Inter-University  Center  for  Astronomy  and  Astrophysics, Pune and the
use of computational facilities at the Institute of Physics, Bhubaneswar.
VPK and JTL acknowledge partial support from the NASA / Space Telescope
Science Institute grant GO-9441. VPK also acknowledges partial support
from the National Science Foundation grant AST-0206197. PK and VPK also
acknowledge partial support from the University of South Carolina
Research Foundation. 

Facilities: \facility{MMT(Blue channel spectrograph)}.

\clearpage


 \begin{figure}
 \figurenum{1}
 \epsscale{0.9}
 \plotone{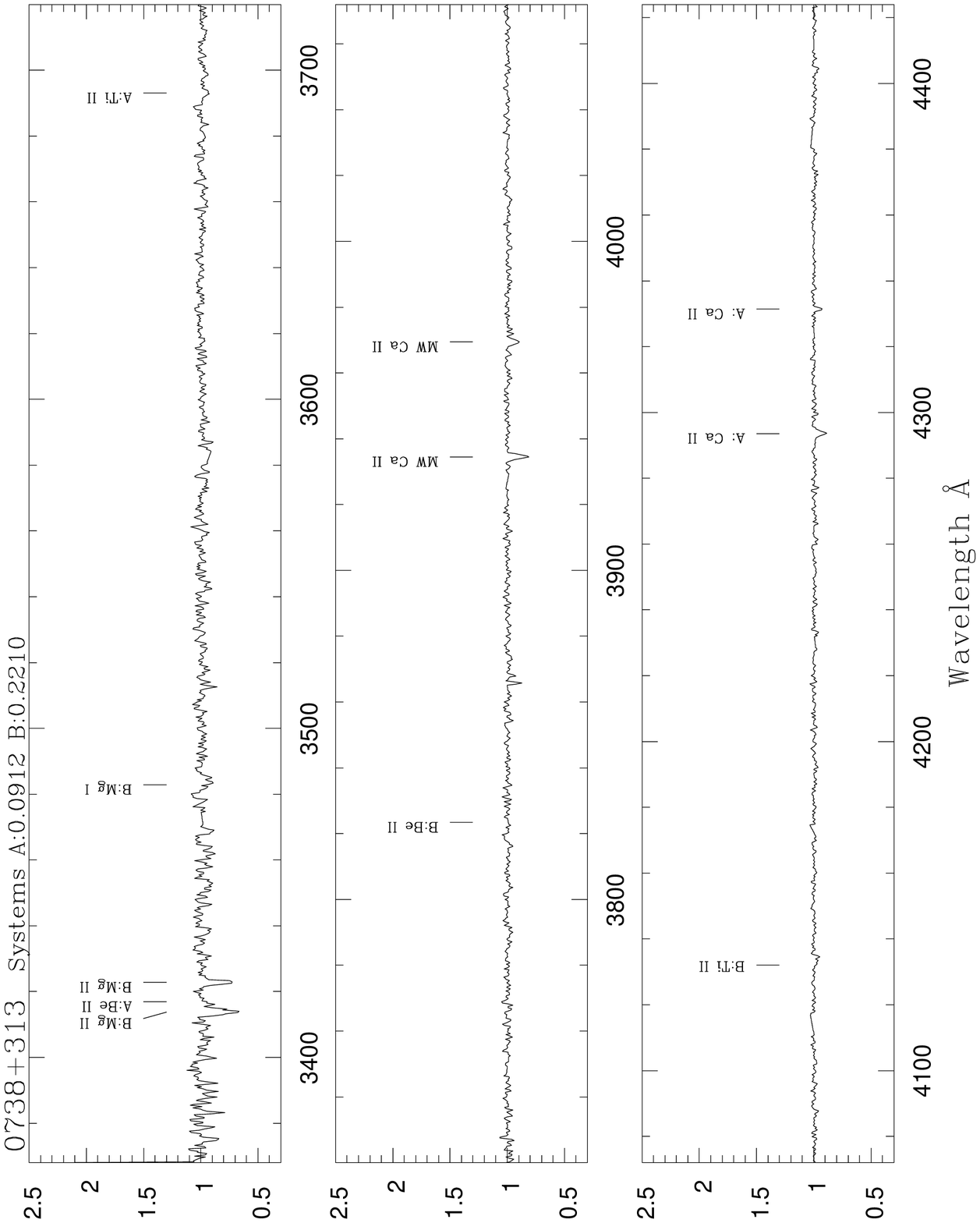}
 \caption{Spectrum of Q0738+313 obtained with the MMT blue 
 channel spectrograph. \label{fig1}}
 \end{figure}
 \clearpage
 \begin{figure}
 \figurenum{2}
 \epsscale{0.9}
 \plotone{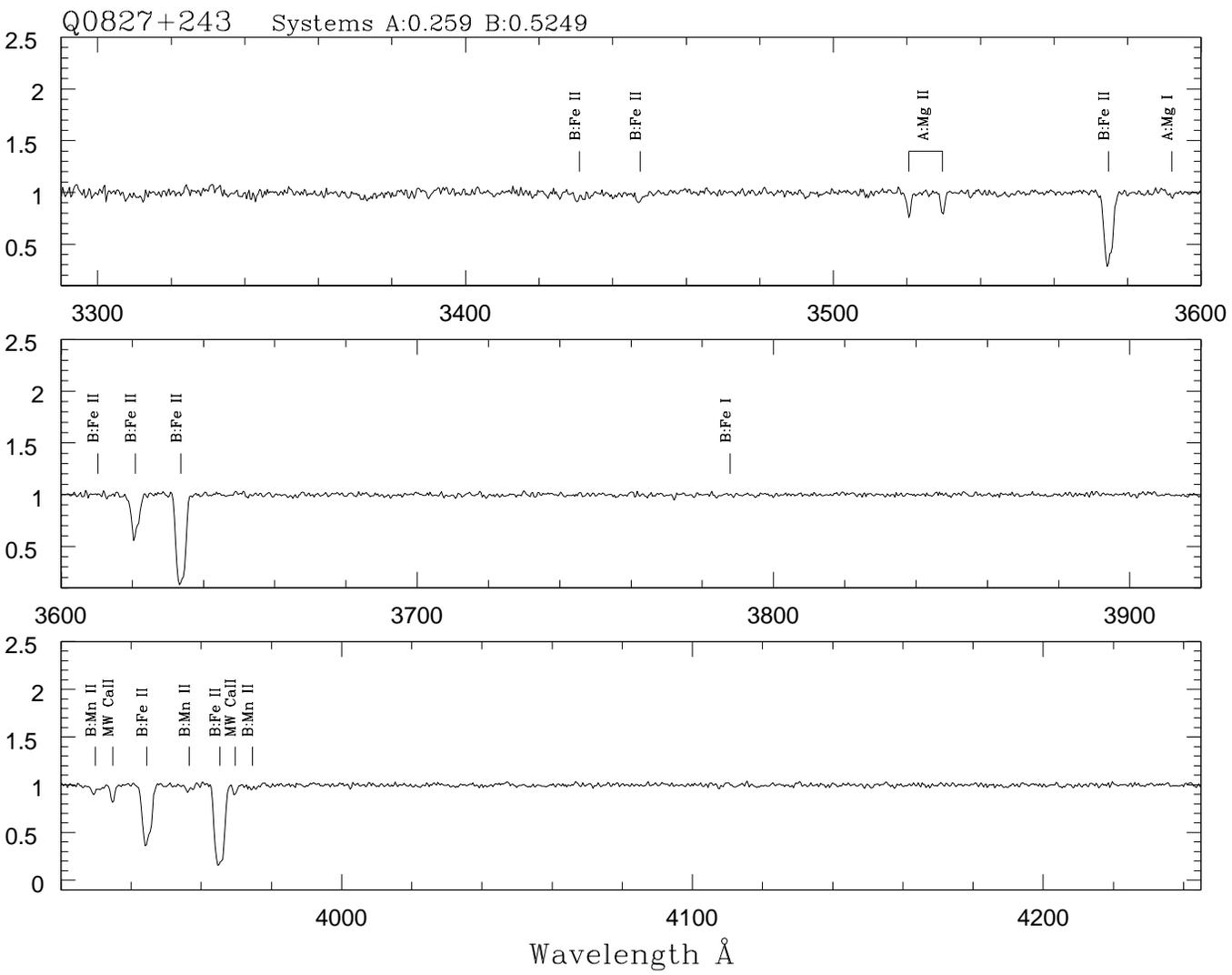}
 \caption{Spectrum of Q0827+243 obtained with the MMT blue 
 channel spectrograph. \label{fig2}}
 \end{figure}
 \clearpage
 
 \begin{figure}
 \figurenum{3}
 \epsscale{0.9}
 \plotone{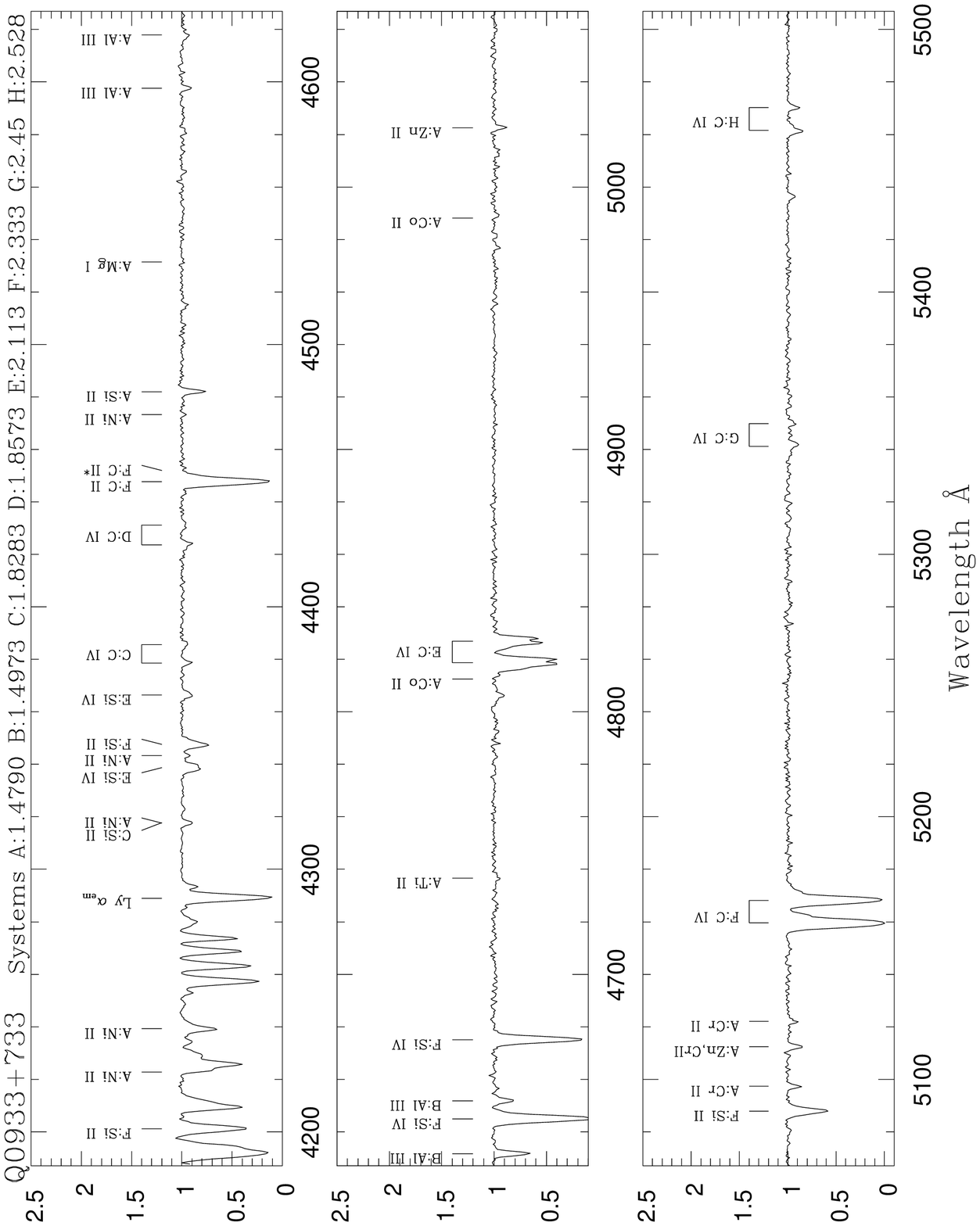}
 \caption{Spectrum of Q0933+733 obtained with the MMT blue 
 channel spectrograph. \label{fig3}}
 \end{figure}
 \clearpage
 \begin{figure}
 \figurenum{4}
 \epsscale{0.85}
 \plotone{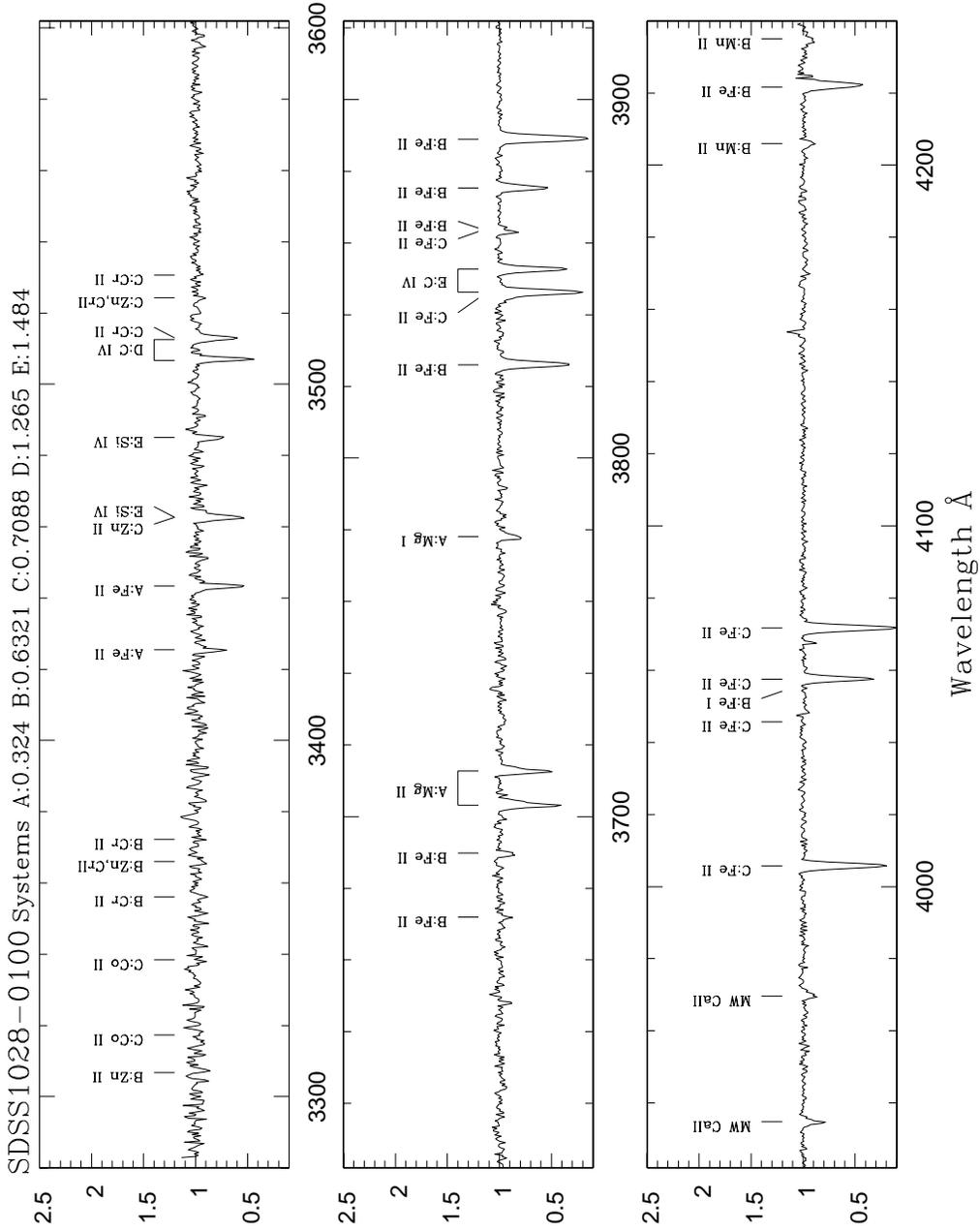}
 \caption{Spectrum of SDSSJ1028-0100 obtained with the MMT blue 
 channel spectrograph. \label{fig4}}
 \end{figure}
 \clearpage
 \begin{figure}
 \figurenum{5}
 \epsscale{0.85}
 \plotone{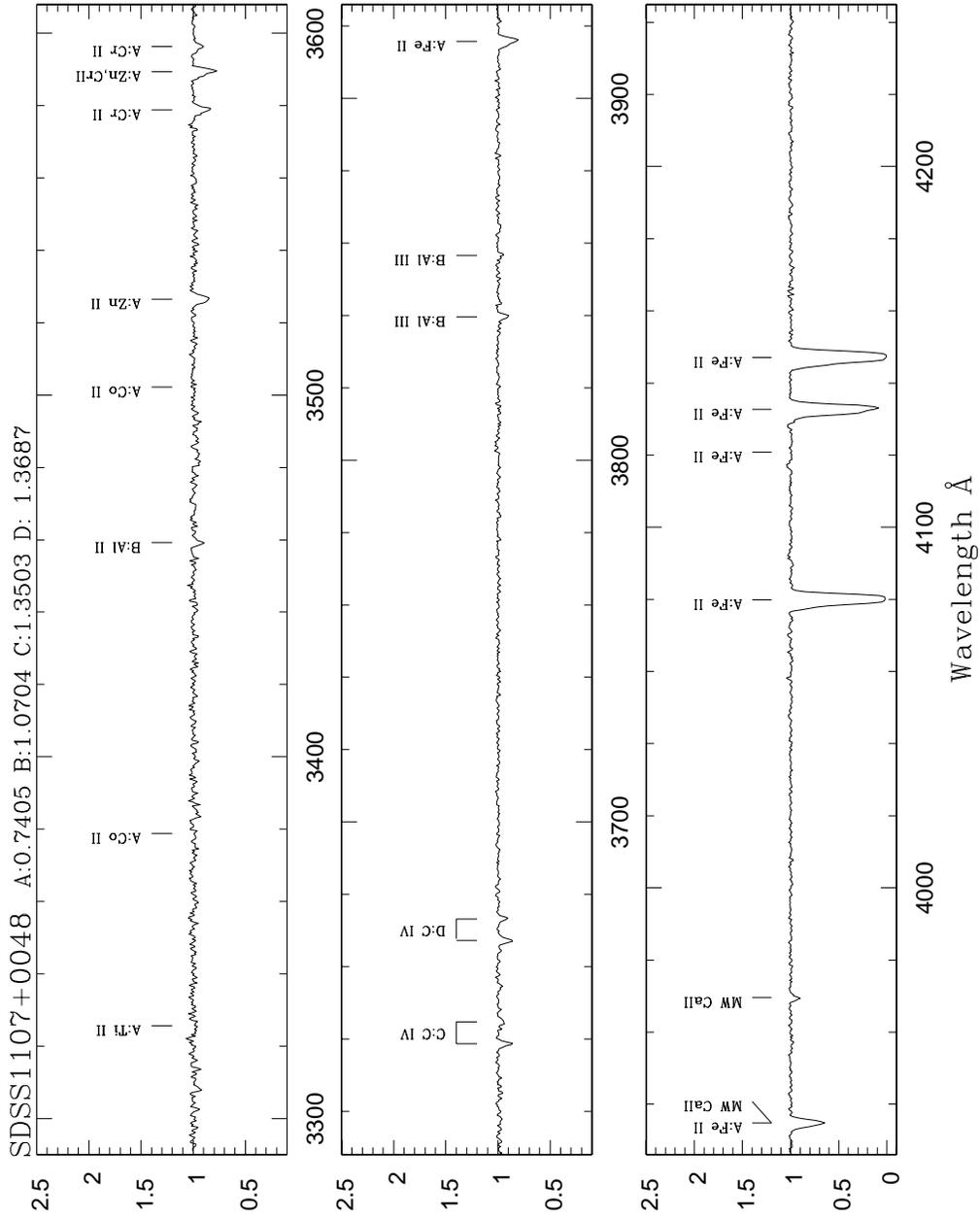}
 \caption{Spectrum of SDSSJ1107+0048 obtained with the MMT 
 blue channel spectrograph. \label{fig5}}
 \end{figure}
 \clearpage
 \begin{figure}
 \figurenum{6}
 \epsscale{0.85}
 \plotone{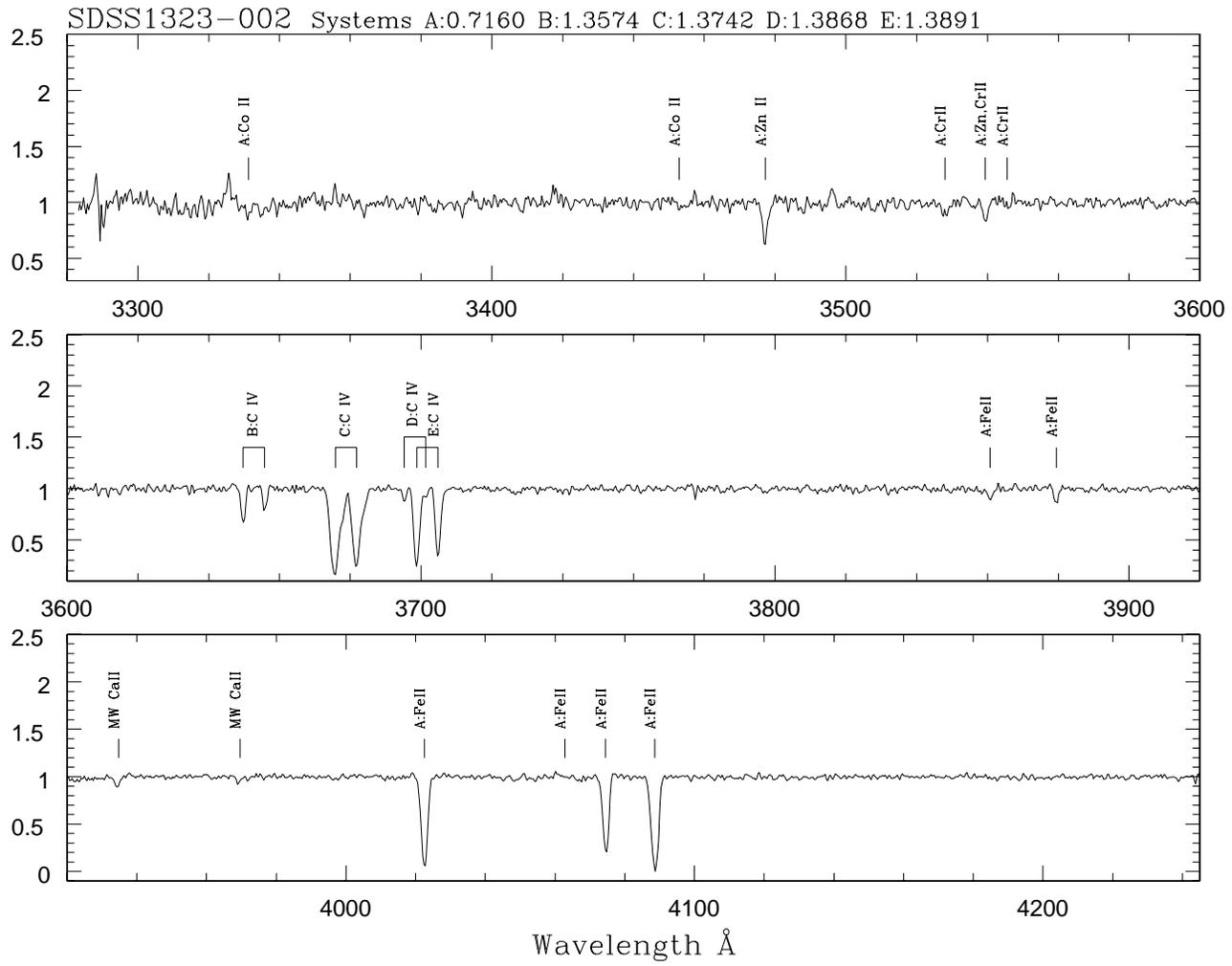}
 \caption{Spectrum of SDSSJ1323-0021 obtained with the MMT blue 
 channel spectrograph. \label{fig6}}
 \end{figure}
 \clearpage
 \begin{figure}
 \figurenum{7}
 \epsscale{0.85}
 \plotone{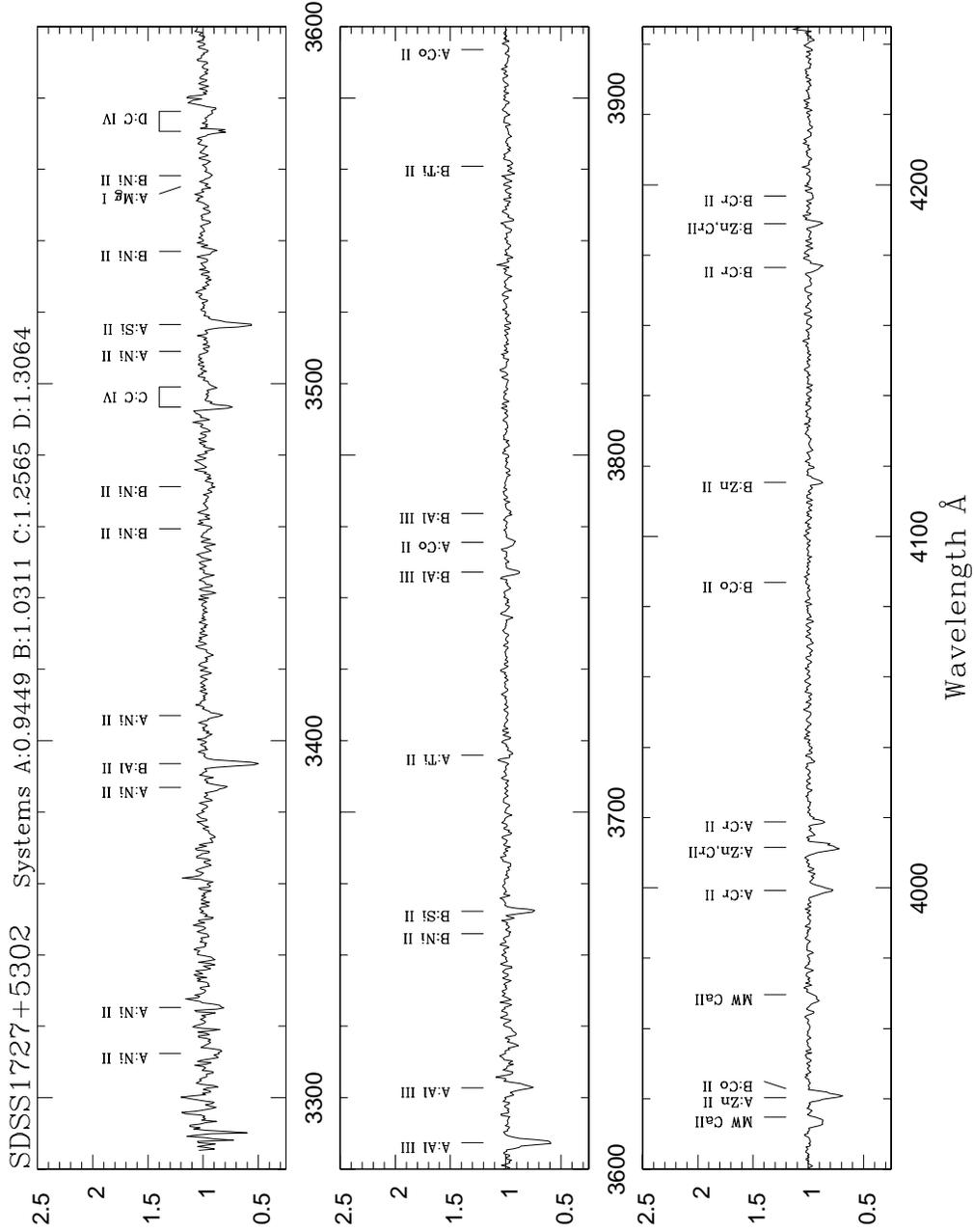}
 \caption{Spectrum of SDSSJ1727+5302 obtained with the MMT blue 
 channel spectrograph. \label{fig7}}
 \end{figure}
 \clearpage
 \begin{figure}
 \figurenum{8}
 \epsscale{0.85}
 \plotone{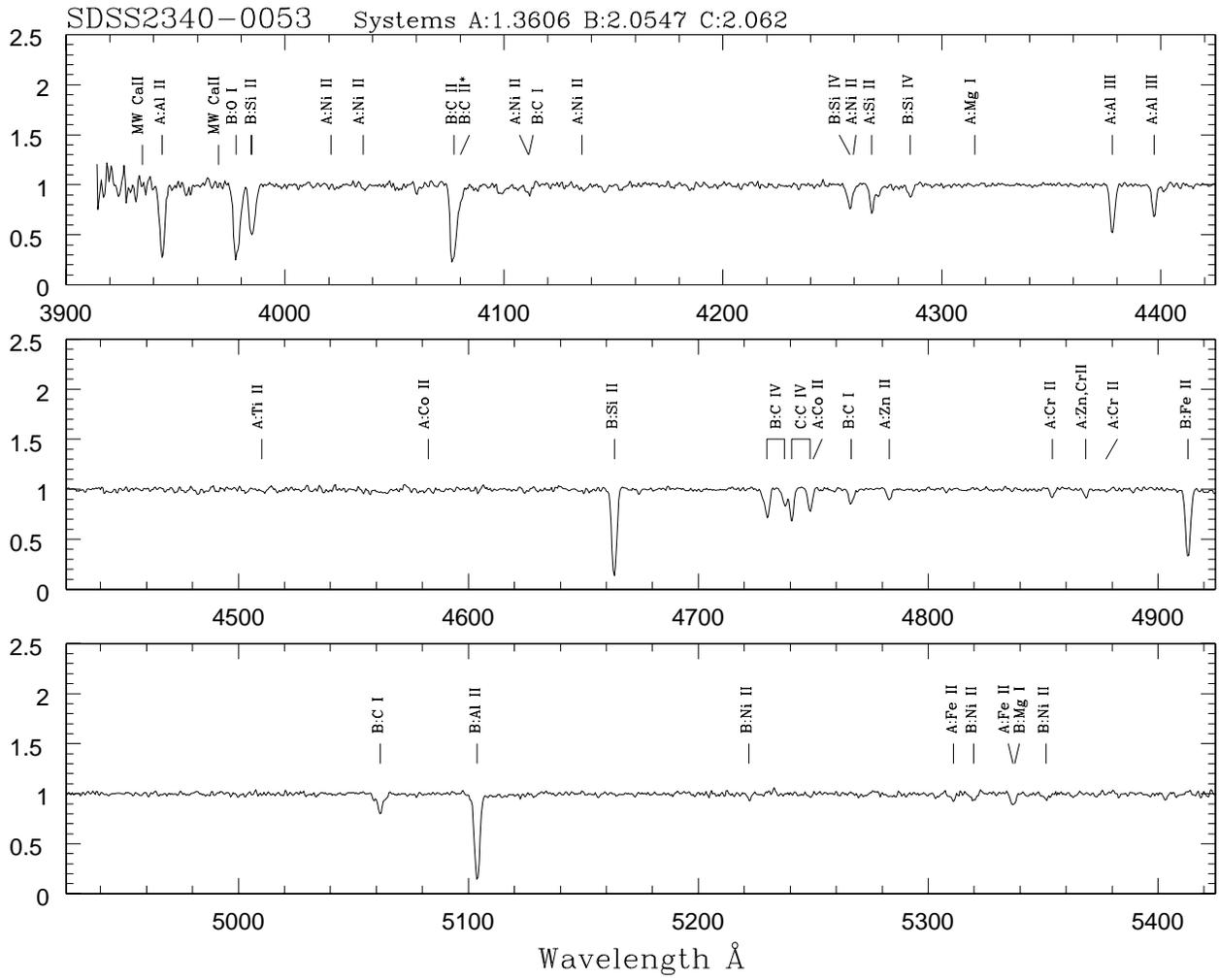}
 \caption{Spectrum of SDSSJ2340-0053 obtained with the MMT blue 
 channel spectrograph. \label{fig8}}
 \end{figure}
 \clearpage 
 
 \begin{figure}
 \figurenum{9}
 \plotone{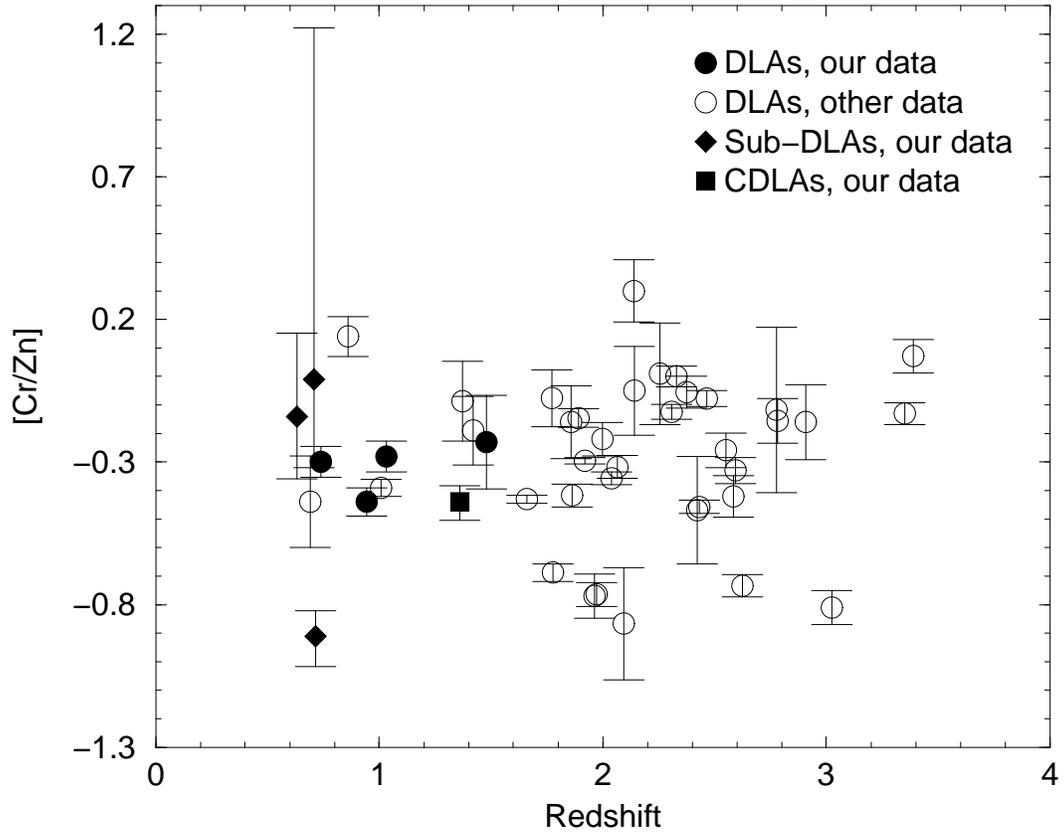}
 \caption{ [Cr/Zn] vs. redshift for DLAs from our MMT survey and other
 data from the literature. Also shown are values for CDLAs and sub-DLAs
 from our data. \label{fig9}}
 \end{figure}
 \clearpage
 \begin{figure}
 \figurenum{10}
 \epsscale{1.0}
 \plottwo{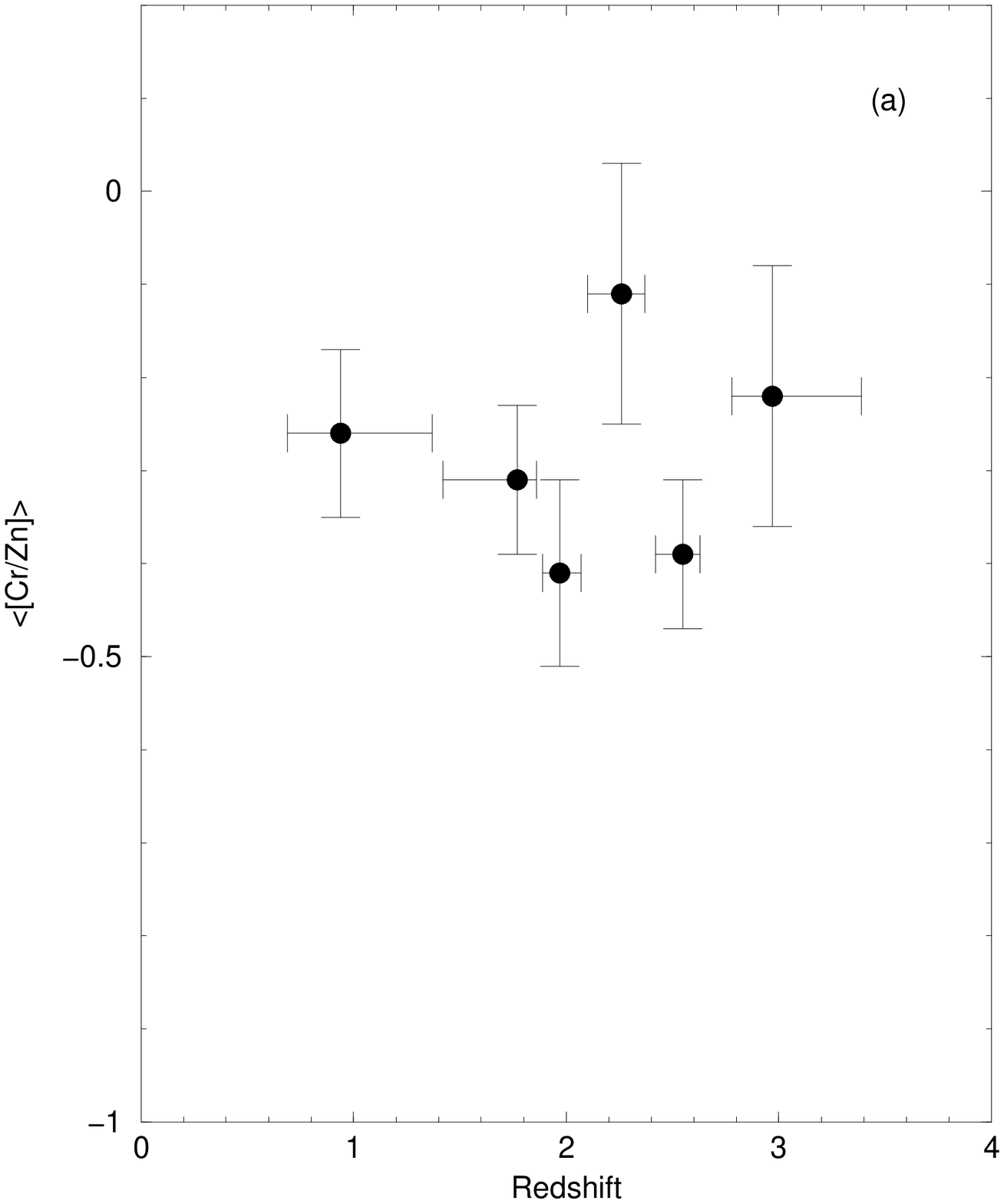}{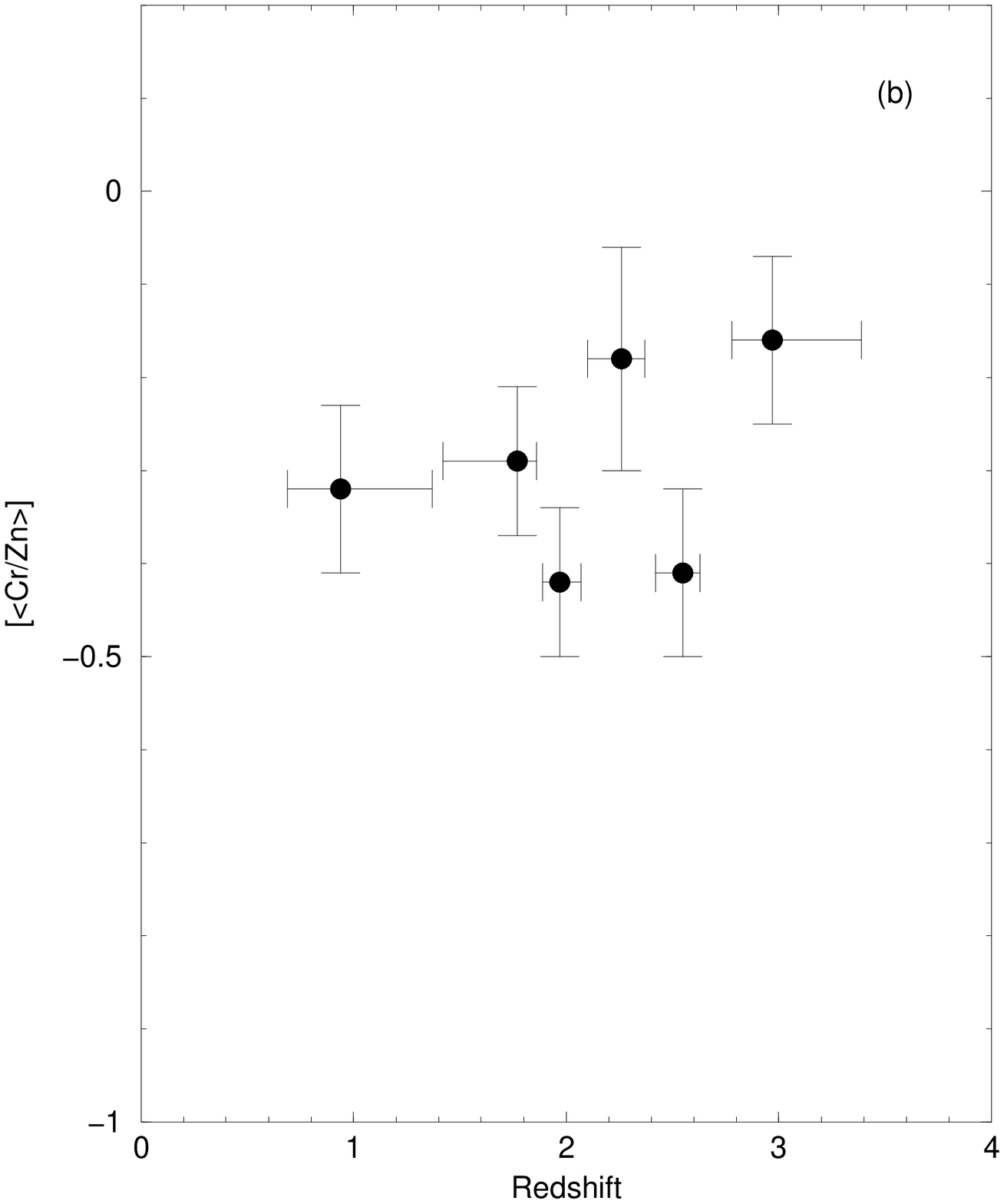}
 \caption{a (Left panel): Binned unweighted mean of logarithmic abundance
 of Cr relative to Zn $\langle$[Cr/Zn]$\rangle$ vs. redshift for DLAs
 from our MMT survey and other data from the literature. b (Right panel):
 Binned logarithm of the $N_{\rm ZnII}$-weighted mean abundance of Cr
 relative to Zn $[ \langle$ Cr/Zn $\rangle ]$ vs. redshift for DLAs from
 our MMT survey and other data from the literature. The zero level in
 both panels denotes the solar level. \label{fig10}}
 \end{figure}
 \clearpage
 \begin{figure}
 \figurenum{11}
 \epsscale{0.6}
 \plotone{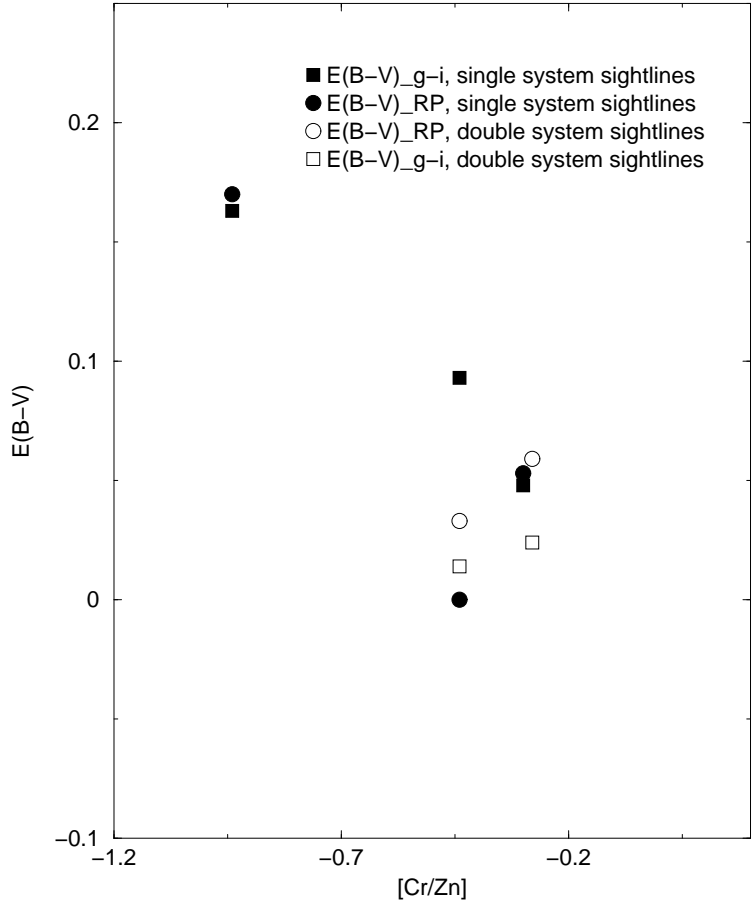}
 \caption{ Reddening $E(B-V)_{RP}$ or $E(B-V)_{g-i}$ relative to SDSS
 composite quasar spectra vs. [Cr/Zn] for the SDSS absorbers from our
 sample. \label{fig11}}
 \end{figure}
 \clearpage




\newpage
\begin{deluxetable}{llrllllcccl}
\tabletypesize{\scriptsize}
\rotate
\tablewidth{470pt}
\tablenum{1}
\tablecaption{Targets Observed}
\tablehead{
\colhead{\bf QSO}&\colhead{Plate}&\colhead{Fiber}
&\colhead{MJD}&\colhead{\bf z$_{\rm
em}$}&\colhead{\bf g}
&\colhead{\bf z$_{\rm DLA}$} &
\colhead{Grating} & \colhead{FWHM}
&\colhead{Time}&\colhead{\bf S/N}\\
\colhead{}&\colhead{}&\colhead{}&\colhead{}&\colhead{} &\colhead{}
&\colhead{}&\colhead{l/mm}  &\colhead{\AA} &\colhead{s} &\colhead{}}
\startdata
Q0738+313 &&&& 0.630\tablenotemark{a}& 16.10& 0.0912 & 832  & 1.10 &
8300\tablenotemark{e}& 15-90\\ 
Q0738+313 &&&& 0.630& 16.10& 0.2210 & 832  & 1.10 & 8300\tablenotemark{e}& 15-90\\ 
Q0827+243 &&&& 0.941\tablenotemark{b} & 17.26& 0.5249 & 832 & 1.10 &
5400\tablenotemark{f} &15-60\\
Q0933+733 &&&& 2.525\tablenotemark{c} & 17.30& 1.4790 & 1200 & 1.33 &
3600\tablenotemark{f} & 50-90 \\
SDSS J102837.01$-$010027.4 &273&286&51957& 1.532\tablenotemark{d} & 18.24& 0.6321 & 832
& 1.04 & 9900\tablenotemark{f} & 20-55\\
SDSS J102837.01$-$010027.4&273&286&51957&1.532\tablenotemark{d}&18.24&0.7088
&832&1.04&9900\tablenotemark{f}&20-55\\
SDSS J110729.03+004811.1&278&378&51900& 1.392\tablenotemark{d} & 17.66& 0.7405 & 832 &
1.04 & 5400\tablenotemark{f} & 10-50\\
SDSS J132323.78$-$002155.2&297&267&51959& 1.390\tablenotemark{d} & 18.49& 0.7160 & 832
& 1.10& 13500\tablenotemark{f} & 25-55\\
SDSS J172739.03+530229.1 &359&42&51821& 1.444\tablenotemark{d} & 18.45& 0.9449 & 832 &
1.10 & 14400\tablenotemark{f} & 35-70\\
SDSS J172739.03+530229.1&359&42&51821&1.444\tablenotemark{d}&18.45& 1.0311
&832&1.10&14400\tablenotemark{f}&35-70\\
SDSS J234023.66$-$005327.0 &385&204&51877& 2.085\tablenotemark{d} & 17.84 & 1.3606 &
1200 & 1.50 & 18900\tablenotemark{e} & 50-90\\
\enddata
\tablenotetext{a}{Veron-Cetty \& Veron (1989)} \tablenotetext{b}{Ulrich \&
Owen (1977)}
\tablenotetext{c}{Steidel \& Sargent (1992)} \tablenotetext{d}{
http://skyserver2.fnal.org}
\tablenotetext{e}{Observed in October, 2002}
 \tablenotetext{f}{Observed in March, 2003} 
\end{deluxetable}
\newpage
\begin{deluxetable}{lllllllllllllll}
\tabletypesize{\scriptsize}
\rotate
\tablenum{2}
\tablecaption{Other data\tablenotemark{a}}
\tablehead{
\colhead{ QSO}&\colhead{ z$_{abs}$}&\colhead{ FeII}&\colhead{
FeII}&\colhead{ FeII}&\colhead{ FeII}&\colhead{ FeII}&\colhead{
MnII}&\colhead{ MnII}&\colhead{ MnII}&\colhead{ MgII}&\colhead{
MgII}&\colhead{ MgI}&\colhead{ CaII}&\colhead{ CaII}\\
\colhead{}&\colhead{}&\colhead{2344}&\colhead{2374}&\colhead{2382}&
\colhead{2586}&\colhead{2600}&\colhead{2576}&\colhead{2594}&
\colhead{2606}&\colhead{2796}&\colhead{2803}&\colhead{2853}&
\colhead{3933}&\colhead{3969}}
\startdata
0738\tablenotemark{b}&0.22&&&&&$<$0.6&&&&0.52&0.50&$<0.2$&&\\
0827\tablenotemark{c}&0.53&&&&&1.90&&&&2.90&2.20&&&\\
0933\tablenotemark{d}&1.48&0.53&0.78&0.57&0.43&0.76
&&&&0.95&1.15&$<0.3$&&\\
&&0.05&0.07&0.05&0.06&0.08&&&&0.08&0.09&&&\\
1028\tablenotemark{e,f} &0.63&0.88&&0.94&0.80
&1.13&$<0.12$&$<0.12$&$<0.12$&1.61&1.36&0.48&0.52&0.39\\
 &&0.15&&0.13&0.18&0.12&&&&0.11&0.11&0.11&0.11&0.11\\
1028\tablenotemark{e,f}& 0.71&0.79&0.61&1.09 &0.98
&0.87&&&0.36&1.22&1.08&0.72&&\\
&& 0.13&0.13&0.19 &0.12&0.11&&&0.08&0.08&0.09&0.11&&\\
1107\tablenotemark{e}& 0.74&1.98 &1.58&2.18&2.02 &2.31
&0.42&0.18&0.24&2.86&2.71&0.84&0.41&0.28\\
&&0.03 &0.03&0.03&0.03 &0.03&0.03&0.03&0.03&0.03&0.03&0.03&0.04&0.04\\
1323\tablenotemark{e}&0.72&1.68&1.17  &1.46&1.43&&&&&2.41&2.13&0.95&&\\
&&0.17&0.17  &0.15& 0.11 &&&&&0.09&0.09&0.07&& \\
1727\tablenotemark{e}& 0.95 &1.65&1.24 &2.01&1.79&2.16
&$<0.08$&$<0.08$&$<0.08$&2.75&2.60&0.94&0.65&0.88  \\
&&0.08&0.06 &0.08&0.08&0.08&&&&0.09&0.09&0.09&0.09&0.20  \\
1727\tablenotemark{e}& 1.03 &0.56&0.39 &0.69&0.54&0.75
&$<0.08$&$<0.08$&$<0.08$&1.01&1.00&0.30&&  \\
&&0.05 &0.06 &0.06&0.08&0.06&&&&0.07&0.07&0.08&& \\
2340\tablenotemark{e,g}&1.36 &1.05&0.76&1.15&1.08&1.21
&$<0.06$&$<0.06$&$<0.06$&1.55&1.58&0.54&$<0.06$&$<0.06$ \\
&&0.03&0.03&0.03&0.03&0.03&&&&0.03&0.03&0.03&& \\
\enddata
\tablenotetext{a}{Columns 3 to 15 list the rest equivalent 
widths in \AA, values in odd rows from 5 to 17 are 1 $\sigma$ errors}
\tablenotetext{b}{Boisse et al. (1992)}
\tablenotetext{c}{Wills 1978 and Ulrich \& Owen (1977)}
\tablenotetext{d}{Steidel \& Sargent, (1992)}
\tablenotetext{e}{York et al. (2000); Stoughton et al. (2002); Aberjazian
et al. (2003)}
\tablenotetext{f}{Petitjean et al. (1998): Si II($\lambda$ 1526):
B(0.51), C(0.46); Al II($\lambda$ 1670): B(0.55), C(0.64); Al
III($\lambda$ 1862): B(0.33), Al
III($\lambda$ 1854): C(0.26)}
\tablenotetext{g}{Additional SDSS data :\\ 
Al II ($\lambda$1670) : 0.92$\pm{0.05}$ ; Si II ($\lambda$ 1808):0.29$\pm{0.03}$ ; Al
III($\lambda$1854): 0.55$\pm{0.04}$ ; Al III($\lambda$1862):0.31$\pm{0.04}$}
\end{deluxetable}
\newpage
\begin{table}
\centerline {Table 3 Atomic data\label{tbl-3}}
\bigskip
\begin{tabular}{|l|l|l|l||l|l|l|l|}
\tableline
\multicolumn{1}{|l|}{\bf Species}&\multicolumn{1}{|l|}{\bf 
$\lambda_{vac}(\AA)$}&\multicolumn{1}{|l|}{\bf f
}&\multicolumn{1}{|l||}{\bf Ref}&\multicolumn{1}{|l|}{\bf
Species}&\multicolumn{1}{|l|}{\bf 
$\lambda_{vac}(\AA)$}&\multicolumn{1}{|l|}{\bf f
}&\multicolumn{1}{|l|}{\bf Ref}\\
\tableline\tableline
  O I & 1302.1685&4.800e-2&a&          Cr II& 2056.2569&1.050e-1& a,c\\ 
  Si II& 1304.3702 &8.630e-2&a&        Cr II& 2062.2361&7.800e-2& a,e\\    
  Si IV & 1393.7602& 5.130e-1&a&       Zn II& 2062.6604&2.560e-1& a,e\\   
  Si IV& 1402.7729&2.540e-1 &a&        Cr II& 2066.161 &5.150e-2& e\\    
  C IV&1548.204&1.899e-1&a &           Fe II& 2249.8768&1.820e-3& f\\    
  C IV& 1550.781& 9.475e-2&a&          Fe II& 2260.7805&2.440e-3& f\\  
  Fe II& 1608.4511 &5.770e-2&a&        Al I & 2263.4644&8.920e-2& a\\ 
 Al II& 1670.7874&1.740e-0&a&          Fe II& 2344.2139&1.140e-1& a,g\\      
 Ni II&1703.4119&5.999e-3&a&           Fe II& 2367.5905&2.160e-5& a\\     
 Ni II& 1709.6042&3.240e-2&a&          Fe II& 2374.4612 &3.130e-2 & a,g\\ 
 Ni II&1741.549 &4.270e-2& a&          Fe II& 2382.7652&3.200e-1& a\\  
 Ni II& 1751.910 &2.270e-2& a&         Fe I & 2484.0211&0.621e-0& a\\ 
 Ni II&1804.473&3.800e-3&b&            Mn II& 2576.877 &3.610e-1& a\\     
 Si II& 1808.0129&2.080e-3&a,c&        Fe II& 2586.6500&6.910e-2& a\\     
 Mg I & 1827.9351&2.420e-2&a&          Mn II& 2594.499 &2.800e-1& a\\     
 Al III&1854.7164&5.590e-1&a&          Fe II& 2600.1729&2.390e-1& a \\    
 Al III&1862.7895&2.780e-1&a&          Mn II& 2606.462 &1.980e-1& a\\     
 Ti II&1910.6123&1.040e-1&a,d&         Mg II& 2796.3543&6.155e-1& a\\    
 Ti II&1910.9538&9.800e-2&a,d&         Mg II& 2803.5315&3.058e-1& a\\     
 Co II&1941.2852&3.403e-2&a&           Mg I & 2852.9631&1.830e-0& a\\      
 Co II&2012.1660&3.680e-2&a&           Be II& 3131.3292 &3.321e-1&a\\     
 Zn II& 2026.1370&4.890e-1& a,e&       Ti II& 3384.7304&3.580e-1& a\\     
 Cr II& 2026.269 &1.300e-3&a,e&        Ca II& 3934.7750&6.267e-1&a\\     
 Mg I & 2026.4768&1.130e-1& a&         Ca II& 3969.5901&3.116e-1&a\\      
                                                                  
\tableline
\end{tabular}
\tablenotetext{References:}{{\bf a:} Morton, (2003); {\bf b:} Welty et al.
(1999); {\bf c:} Bergeson \& Lawler (1993a); {\bf d:} Wiese, Fedchak \&
Lawler (1996); {\bf e:} Bergeson \& Lawler (1993b); {\bf f:} Bergeson,
Mullman \& Lawler (1994)
{\bf g:} Bergeson et al. (1996) } 
\end{table}
\clearpage


\clearpage

\setlength{\hoffset}{-0.6in}
\setlength{\voffset}{-0.6in}
\begin{table}
\centerline{\qquad Table 4 Equivalent Widths}
\bigskip

\begin{tabular}{|l|l|l|l|r|l||l|l|l|l|r|l|}
\tableline
\multicolumn{1}{|l|}{\bf QSO}&\multicolumn{1}{|l|}{\bf
$z_{abs}$}&\multicolumn{1}{|l|}{\bf Species}&\multicolumn{1}{|l|}{\bf
$\lambda_{\rm rest}$}&\multicolumn{1}{|l|}{\bf
$\Delta$z$_{abs}^a$}&\multicolumn{1}{|l||}{\bf W$_{\rm
rest}$}&\multicolumn{1}{|l|}{\bf QSO}&\multicolumn{1}{|l|}{\bf
$z_{abs}$}&\multicolumn{1}{|l|}{\bf Species}&\multicolumn{1}{|l|}{\bf
$\lambda_{\rm rest}$}&\multicolumn{1}{|l|}{\bf
$\Delta$z$_{abs}^a$}&\multicolumn{1}{|l|}{\bf W$_{\rm rest}$}\\
&&&\AA&&\AA&&&&\AA&&\AA\\
\tableline\tableline
0738&0.0912 & Be II&3131&&$<0.1$&                   && Fe II &  2249 &-14.4 & 0.08$\pm{0.01}$\\         
A&&Ti II &  3384 & -5.0 &0.13$\pm{0.02}$&           && Fe II &  2260 &-12.5 & 0.11$\pm{0.01}$\\         
&& Ca II&3934&5.8 &0.16$\pm{0.02}$&                 && Fe II &  2344 & 5.3 & 0.78$\pm{0.03}$\\  
&&Ca II &  3969 & -0.8 &0.08$\pm{0.01}$ &           && Fe II & 2367 & &$<$0.13\\    
0738&0.2210 &Mg II&2796&   -7.6&0.61$\pm{0.04}$&     && Fe II &  2374 & -0.5 & 0.44$\pm{0.01}$ \\      
B&& Mg II &  2803 & -7.3 &0.38$\pm{0.02}$&            && Fe II &  2382 &  7.5 & 1.05$\pm{0.02}$\\     
&& Mg I  &  2853 & 14.9 & 0.24$\pm{0.04}$&            && Fe I  &  2484 && $<$0.02  \\                 
&&Be II & 3131 && $<$0.03&                            && Mn II &  2576 &  4.6 & 0.11$\pm{0.01}$\\     
&&Ti II & 3384 && $<$0.04&                            && Fe II &  2586 & 13.7 & 0.70$\pm{0.02}$\\      
0827 &0.5249 &Fe II& 2249&3.1& 0.13$\pm{0.01}$&       && Mn II &  2594 &  4.4 & 0.12$\pm{0.03}$\\     
B&& Fe II &  2260 & -4.2 & 0.14$\pm{0.01}$&          1028&0.7088& Co II & 1941 &&$<$0.06\\     
&& Fe II &  2344 & -5.9 & 1.11$\pm{0.01}$&           C&& Co II & 2012 &&$<$0.06\\     
&& Fe II &  2367 &  & $<$0.03&                       && Zn/Mg$^{b,d}$ & 2026&&$<$0.48\\     
&& Fe II &  2374 & -2.9 & 0.67$\pm{0.02}$&           && Cr II$^d$ & 2056 &&$<$0.38\\    
&& Fe II &  2382 &  4.0 & 1.69$\pm{0.02}$&            &&Zn/Cr$^c$ &  2062 & & 0.05$\pm{0.02}$\\        
&&Fe I& 2484&& $<$0.02&                               && Cr II &  2066 & 45.9 & 0.07$\pm{0.02}$\\    
&& Mn II &  2576 &  0.7 & 0.20$\pm{0.01}$&            && Fe II &2249&&$<$0.91\\     
&& Fe II &  2586 & -0.1 & 1.19$\pm{0.01}$&            && Fe II$^d$&2260&  3.8&0.13$\pm{0.01}$\\     
&& Mn II &  2594 &  0.4 & 0.10$\pm{0.01}$&            && Fe II &  2344 & -3.3 & 0.79$\pm{0.01}$\\     
&& Fe II &  2600 &  4.9 & 1.79$\pm{0.02}$&            && Fe II &  2367 & & $<0.03$\\     
&& Mn II & 2606 &-4.2& 0.10$\pm0.02$&                 && Fe II &  2374 &  0.5 & 0.60$\pm{0.01}$\\     
0933& 1.4790 &Ni II &  1703&&$<$1.06 &                && Fe II &  2382 & -1.1 & 0.96$\pm{0.01}$\\      
A&&Ni II &  1709&&$<$1.40 &                           1107&0.7405&Ti II & 1910 & &$<$0.02\\      
&&Ni II &  1741 & 18.6 & 0.08$\pm{0.01}$&             A&& Co II & 1941 &&$<$0.02\\    
&& Ni II &  1751 & 15.2 & 0.06$\pm{0.01}$&            && Co II & 2012 &&$<$0.02\\   
&&Ni II &  1804 & -3.4 & 0.02$\pm{0.01}$&             &&Zn/Mg$^b$ & 2026 && 0.20$\pm{0.01}$\\     
&& Si II &  1808 & -2.3 & 0.16$\pm{0.01}$&            && Cr II &  2056 & -4.3 & 0.19$\pm{0.01}$\\    
&& Mg I&1827&& $<$0.01&                               && Zn/Cr$^c$ &  2062 & & 0.25$\pm{0.01}$\\       
&& Al III&  1854 & -12.8& 0.06$\pm{0.01}$&            && Cr II &  2066 & 13.0 & 0.12$\pm0.01{}$\\     
&& Al III$^d$& 1862 & -1.2 & 0.08$\pm{0.01}$&         && Fe II & 2249& -2.0 & 0.26$\pm{0.01}$\\     
&& Ti II &  1910 & -25.6& 0.07$\pm{0.01}$&            && Fe II$^d$ & 2260 &-2.7 & 0.43$\pm{0.01}$\\    
&& Co II & 1941 && $<$0.02&                          && Fe II &  2344 & -3.1 & 1.95$\pm{0.01}$\\     
&& Co II & 2012 && $<$0.02 &                         && Fe II &  2367 & & $<0.02$\\     
&& Zn/Mg$^b$&  2026 &  & 0.08$\pm{0.01}$&             && Fe II &  2374 &  0.4 & 1.56$\pm{0.01}$\\     
&& Cr II &  2056 & -7.3 & 0.08$\pm{0.01}$&            && Fe II &  2382 & -4.1 & 2.15$\pm{0.01}$\\     
&& Zn/Cr$^c$ &  2062 & & 0.11$\pm{0.01}$&            1323&0.7160& Co II & 1941 &&$<$0.04\\     
&& Cr II$^d$ &  2066 & 5.0  & 0.06$\pm{0.01}$&      A&& Co II & 2012 &&$<$0.04\\     
1028& 0.6321& Zn/Mg$^b$&  2026 &  & 0.08$\pm{0.02}$&  && Zn/Mg$^b$&2026 && 0.38$\pm{0.02}$\\     
B&& Cr II &  2056 & -8.2 & 0.09$\pm{0.02}$&           && Cr II $^d$&2056 &-16.1 & 0.18$\pm{0.02}$\\    
&& Zn/Cr$^c$ &  2062 & & 0.09$\pm{0.02}$&             && Zn/Cr$^c$ & 2062 & & 0.16$\pm{0.02}$\\     
&& Cr II & 2066 && $<$0.06&                           && Cr II & 2066 &&$<$0.04\\                      
                                                      
\tableline  
\end{tabular}
\end{table}
\clearpage

\begin{table}
\centerline{Equivalent Widths (cont'd.)\label{tbl-4}}
\bigskip

\begin{tabular}{|l|l|l|l|r|l||l|l|l|l|r|l|}
\tableline
\multicolumn{1}{|l|}{\bf QSO}&\multicolumn{1}{|l|}{\bf
$z_{abs}$}&\multicolumn{1}{|l|}{\bf Species}&\multicolumn{1}{|l|}{\bf
$\lambda_{\rm rest}$}&\multicolumn{1}{|l|}{\bf
$\Delta$z$_{abs}^a$}&\multicolumn{1}{|l||}{\bf W$_{\rm
rest}$}&\multicolumn{1}{|l|}{\bf QSO}&\multicolumn{1}{|l|}{\bf
$z_{abs}$}&\multicolumn{1}{|l|}{\bf Species}&\multicolumn{1}{|l|}{\bf
$\lambda_{\rm rest}$}&\multicolumn{1}{|l|}{\bf
$\Delta$z$_{abs}^a$}&\multicolumn{1}{|l|}{\bf W$_{\rm rest}$}\\
&&&\AA&&\AA&&&&\AA&&\AA\\
\tableline\tableline
  && Fe II & 2249 &-10.1&0.11$\pm{0.02}$&               && Al III& 1854 &  7.2 & 0.09$\pm{0.01}$\\                  
  && Fe II & 2260 &  6.6 & 0.13$\pm{0.01}$&             && Al III& 1862 & 16.0 & 0.04$\pm{0.01}$\\                  
  && Fe II &  2344 &  5.6 & 1.26$\pm{0.01}$&            && Ti II & 1910 &&$<$0.02\\
  && Fe II &  2367 &&$<0.02$&                           && Co II & 1941 &&$<$0.02\\              
  && Fe II &  2374 & 10.1 & 0.95$\pm{0.01}$&              && Co II & 2012 &&$<$0.01\\ 
 && Fe II &  2382 &  4.0 & 1.47$\pm{0.01}$&               && Zn/Mg$^b$&  2026 &  & 0.12$\pm{0.01}$\\  
  1727& 0.9449 & Ni II &  1703 &-32.4 & 0.20$\pm{0.04}$&  && Cr II & 2056 &  4.9 & 0.12$\pm{0.01}$\\     
 A&& Ni II &  1709& 24.3 & 0.18$\pm{0.02}$ &              && Zn/Cr$^c$ &  2062 & & 0.08$\pm{0.01}$\\      
  && Ni II &  1741 &-22.4 & 0.28$\pm{0.03}$&              && Cr II & 2066 &  7.3 & 0.04$\pm{0.01}$\\     
  && Ni II &  1751 &-12.1 & 0.14$\pm{0.02}$&              2340 & 1.3606 & Al II &  1670 &-11.2 & 0.89$\pm{0.03}$\\  
  && Ni II &  1804 &-30.8 & 0.02$\pm{0.01}$&              A&& Ni II & 1703 &&$<0.02$\\   
  && Si II &  1808 & 7.8 & 0.45$\pm{0.02}$ &              && Ni II &  1709 &40.4 & 0.05$\pm{0.01}$\\   
  && Mg I  &  1827 &&$<$0.04               &              && Ni II &  1741 & 10.3 & 0.12$\pm{0.01}$\\   
  && Al III&  1854 &  9.9 & 0.42$\pm{0.02}$&              && Ni II &  1751 &  8.6 & 0.10$\pm{0.01}$\\   
  && Al III&  1862 & -9.3 & 0.26$\pm{0.02}$&              && Ni II & 1804 &&$<0.29$\\   
  && Ti II &  1910 & 15.5 & 0.06$\pm{0.01}$&              && Si II$^d$ &  1808 & 3.05 & 0.29$\pm{0.01}$\\   
  && Co II$^d$ &  1941 & -6.9 & 0.08$\pm{0.01}$&          && Mg I & 1827 &&$<$ 0.02\\   
  && Co II & 2012&&$<$0.03                 &              && Al III&  1854 &-17.6 & 0.52$\pm{0.01}$\\   
  && Zn/Mg$^b$&  2026 & & 0.32$\pm{0.01}$  &              && Al III&  1862 & 2.3 & 0.38$\pm{0.01}$\\   
  && Cr II &  2056 & 21.1 & 0.34$\pm{0.02}$&              && Ti II & 1910 &&$<$0.02\\   
  && Zn/Cr$^c$ &  2062 & & 0.33$\pm{0.02}$ &              && Co II & 1941 &&$<$0.2\\   
  && Cr II &  2066 & 19.0 & 0.13$\pm{0.01}$&              && Co II & 2012 &&$<$0.20\\    
1727& 1.0311 & Al II &  1670 &  4.6 & 0.41$\pm{0.02}$&    && Zn/Mg$^b$&  2026 & & 0.10$\pm{0.01}$\\   
B&& Ni II & 1703 && 0.05$\pm0.02$&                        && Cr II &  2056 &  0.5 & 0.07$\pm{0.01}$\\   
&& Ni II &  1709 &-32.9 & 0.10$\pm{0.03}$&                && Zn/Cr$^c$ &  2062 & & 0.07$\pm{0.01}$\\   
&& Ni II &  1741 & -8.7 & 0.08$\pm{0.01}$&                && Cr II & 2066 &&$<$0.02 \\   
&& Ni II &  1751 & 16.1 & 0.08$\pm{0.02}$&                && Fe II &  2249 &-22.9 & 0.07$\pm{0.02}$\\   
&& Si II & 1808 & 1.9 & 0.19$\pm{0.01}$&                 && Fe II$^d$ &  2260 & 7.3  & 0.13$\pm{0.01}$\\                   
\tableline
\end{tabular}
${\bf a:}$ In units of 10$^{-5}$ \\
${\bf b:}$ Zn II (2026) is blended with Mg I  (2026)\\
${\bf c:}$ Zn II (2062) is blended with Cr II (2062)\\
${\bf d:}$ Blend\\
\end{table}

\clearpage

\begin{table}
\centerline{Table 5 Column Densities\label{tbl-5} }
\bigskip
\begin{tabular}{|l|l|l|l|l||l|l|l|l|l|}
\tableline
\multicolumn{1}{|l|}{\bf QSO}&\multicolumn{1}{|l|}{\bf z$_{\rm
abs}$}&\multicolumn{1}{|l|}{\bf Species}&\multicolumn{1}{|l|}{\bf log N
}&\multicolumn{1}{|l||}{\bf $b_{eff}$ }&\multicolumn{1}{|l|}{\bf
QSO}&\multicolumn{1}{|l|}{\bf z$_{\rm abs}$}&\multicolumn{1}{|l|}{\bf
Species}&\multicolumn{1}{|l|}{\bf log N }&\multicolumn{1}{|l|}{\bf
$b_{eff}$ }\\
&&& cm$^{-2}$&km s$^{-1}$&&&&cm$^{-2}$&km s$^{-1}$\\
\tableline\tableline
0738 & 0.0912 & H I$^a$&21.18$^{0.05}_{0.06}$         && 1028 & 0.7088 & H I$^c$& 20.01$^{0.12}_{0.17}$&\\   
A&& Be II & $<$12.70 & 10.0                           &C&&Mg I$^b$ &$13.13^{0.19}_{0.17}$ & 30.0\\   
&& Ca II & $12.31^{0.03}_{0.03}$ & $42.3\pm$ 6.9      &&& Al II$^b$& 14.19&30.0\\  
&& Ti II & $12.53^{0.14}_{0.06}$ & 42.3               &&& Al III$^b$&13.34&30.0\\  
0738 & 0.2210 & H I$^a$ & 20.90$^{0.07}_{0.09}$&      & && Si II$^b$ &14.81 & 30.0\\     
B&& Be II & $<$12.04&10.0                             &&& Cr II & $13.23^{0.09}_{0.11}$ & 30.0\\     
&& Mg I  & 12.00$^{0.09}_{0.12}$& 37.8                 &&& Mn II$^b$ & 13.33 $^{0.17}_{0.19}$ & 30.0\\       
&& Mg II & $>$13.30$^{0.03}_{0.03}$&37.8$\pm$6.3      &&& Fe II &$14.94^{0.03}_{0.03}$ & 30.0$\pm$0.7\\     
&& Ti II & $<$12.11&10.0                               &&& Co II & $<$13.74 & 10.0\\     
0827 & 0.5249 & H I$^a$&20.30$^{0.04}_{0.05}$&           &&& Zn II & $12.22^{0.29}_{1.23}$ & 30.0\\     
B&& Mn II & $12.83^{0.19}_{0.34}$ & 38.2,28.7         & 1107 & 0.7405 & H I$^c$& 20.98$^{0.12}_{0.17}$&\\    
&& Fe I  & $<$11.72& 10.0                             &A&&Mg I &$12.73^{0.27}_{0.82}$& 80.7$\pm$20.4\\     
&& Fe II & $14.59^{0.02}_{0.02}$ & 38.2$\pm1.9$       &&& Ti II & $<$12.82 & 10.0\\     
&& Fe II & $14.23^{0.03}_{0.03}$ & 28.7$\pm1.1$       &&& Cr II &$13.75^{0.03}_{0.03}$ & 80.7$\pm$20.4\\     
0933 & 1.4790 & H I$^a$ &21.62$^{0.08}_{0.09}$&        &&& Mn II$^b$ & 13.34$^{0.04}_{0.07}$&80.7\\    
A&& Mg I  & $<$12.75 & 28.6                           &&& Fe II & $15.52^{0.06}_{0.06}$ &89.5  $\pm$3.5\\     
&& Al III&$12.67^{0.04}_{0.05}$ &  31.3$\pm$14.5      &&& Co II & $<$13.18 & 10.0\\       
&& Si II & $15.52^{0.03}_{0.03}$ & 28.6$\pm$ 7.5       &&& Zn II & $13.03^{0.05}_{0.05}$ & 80.7$\pm$20.4\\      
&& Ti II & $12.85^{0.21}_{0.09}$ & 28.6                & 1323 & 0.7160 & H I$^c$& 20.21$^{0.21}_{0.18}$&\\ 
&& Cr II & $13.46^{0.08}_{0.09}$ & 28.6               &A&&Mg I$^b$ &$13.15^{0.07}_{0.08}$  & 47.0\\   
&& Fe II$^b$ & 14.34&28.6                             &&& Cr II & $13.33^{0.08}_{0.10}$ & 47.0\\      
&& Co II & $<$13.30 & 10.0                            &&& Fe II & $15.11^{0.03}_{0.03}$ & 47.0$\pm$1.0\\     
&& Ni II & $13.91^{0.01}_{0.01}$ & 37.2$\pm$18.1      &&& Co II & $<$13.60 & 10.0\\     
&& Zn II & $12.67^{0.12}_{0.16}$ & 28.6               &&& Zn II & $13.25^{0.04}_{0.04}$ & 47.0\\     
1028 & 0.6321 & H I$^c$& 19.90$^{0.12}_{0.17}$&       & 1727 & 0.9449& H I$^d$ & 21.16$^{0.11}_{0.16}$ &\\     
B&&Mg I$^b$& $12.80^{0.20}_{0.28}$ & 26.7             &A&& Mg I  & $12.90^{0.23}_{0.50}$ & 69.2$\pm$4.2\\     
&& Al II$^b$& 14.06&26.7                               &&& Al III& $13.49^{0.02}_{0.02}$ & 69.3$\pm$4.9\\     
&& Al III$^b$&13.85&26.7                                 &&& Si II & $15.94^{0.02}_{0.02}$ & 61.4$\pm$4.9\\    
&& Si II$^b$ & 15.28 &26.7                              &&& Ti II &$12.90^{0.09}_{0.12}$ & 37.2$\pm$44.6\\     
&& Cr II & $13.34^{0.11}_{0.15}$ & 26.7                &&& Cr II & $13.85^{0.02}_{0.02}$ & 69.2$\pm$4.2\\     
&& Mn II & $12.83^{0.19}_{0.33}$ & 64.6$\pm$ 18.9     &&& Fe II$^b$ & $ 15.38^{0.13}_{0.15}$ & 69.2$\pm$4.2\\     
&& Fe I  & $<11.86$ &10.0                             &&& Co II & $<$13.29 & 10.0\\     
&& Fe II & $15.07^{0.04}_{0.05}$ & 26.7$\pm$ 6.7       &&& Ni II & $14.27^{0.07}_{0.08}$ & 73.1$\pm$23.9\\       
&& Zn II & $12.46^{0.16}_{0.27}$ & 26.7               &&& Zn II & $13.27^{0.04}_{0.05}$ & 69.2$\pm$4.2\\                            
\tableline
\end{tabular}
\end{table}
\clearpage
\begin{table}
\centerline{Column Densities(cont'd)}
\bigskip
\begin{tabular}{|l|l|l|l|l||l|l|l|l|l|}
\tableline
\multicolumn{1}{|l|}{\bf QSO}&\multicolumn{1}{|l|}{\bf z$_{\rm
abs}$}&\multicolumn{1}{|l|}{\bf Species}&\multicolumn{1}{|l|}{\bf log N
}&\multicolumn{1}{|l||}{\bf $b_{eff}$ }&\multicolumn{1}{|l|}{\bf
QSO}&\multicolumn{1}{|l|}{\bf z$_{\rm abs}$}&\multicolumn{1}{|l|}{\bf
Species}&\multicolumn{1}{|l|}{\bf log N }&\multicolumn{1}{|l|}{\bf
$b_{eff}$ }\\
&&& cm$^{-2}$&km s$^{-1}$&&&&cm$^{-2}$&km s$^{-1}$\\
\tableline\tableline

1727 & 1.0311 & H I$^d$ & 21.41$^{0.11}_{0.15}$       &&2340 & 1.3606 & Mg I  & $12.83^{0.11}_{0.15}$ & 33.7$\pm$8.2\\ 
B&& Mg I$^b$  & $12.45^{0.01}_{0.02}$ & 32.4$\pm$8.3  &A&& Al I  & $<12.49$ & 10.0\\ 
&& Al II & $13.15^{0.02}_{0.02}$ & 44.4$\pm$3.6       &&& Al II & $13.53^{0.02}_{0.02}$ & 84.3$\pm$4.9\\ 
&& Al III& $12.75^{0.05}_{0.06}$ & 33.2$\pm$17.7      &&& Al III& $13.63^{0.01}_{0.01}$ & 65.1$\pm$2.4\\ 
&& Si II & $15.60^{0.03}_{0.03}$ & 39.9$\pm$9.0       &&& Si II & $15.70^{0.02}_{0.02}$ & 49.8$\pm$6.5\\ 
&& Ti II & $<$12.92 & 10.0                            &&& Ti II & $<$12.71 & 10.0\\ 
&& Cr II & $13.39^{0.03}_{0.03}$ & 32.4$\pm$8.3       &&& Cr II & $13.20^{0.04}_{0.04}$& 33.7$\pm$8.2\\ 
&& Fe II$^b$ & 14.54&32.4                          &&& Fe II & 14.94$^{0.11}_{0.15}$ &53.2$\pm $43.5\\ 
&& Co II & $<$13.00 & 10.0                            &&& Co II & $<$13.30 & 10.0\\ 
&& Ni II & $13.91^{0.07}_{0.08}$ & 39.4$\pm$23.0      &&& Ni II & $<$13.74 & 10.0\\ 
&& Zn II & $12.65^{0.04}_{0.05}$ & 32.4$\pm$8.3       &&& Zn II & $12.62 ^{0.04}_{0.05}$ & 33.7$\pm$8.2\\ 
\tableline                                                                                                     
\end{tabular}
\tablenotetext{a :} {Rao \& Turnshek, (2000)}
\tablenotetext{b :} {Calculated from equivalent width in Table 2 }
\tablenotetext{c :} {HST GO Project No. 9382, P.I. Rao, S.}
\tablenotetext{d :} {Turnshek et al. (2004)}
\end{table}
\clearpage
\begin{deluxetable}{lrrrrrrrrrr}
\tabletypesize{\footnotesize}
\rotate
\tablenum{6}
\tablecaption{Relative Abundances}
\tablehead{
\colhead{QSO}&\colhead{$z_{abs}$
}&\colhead{[Cr/Zn]}&\colhead{[Fe/Zn]}&\colhead{[Ni/Zn]}&\colhead{[Si/Zn]}&\colhead{[Ti/Zn]}&\colhead{[Co/Zn]}&\colhead{[Mn/Zn]}&\colhead{[Al/Zn]}&\colhead{[Si/Fe]}}
\startdata
0827 & 0.53 &...        &...          &...      &...        &...      &...     &...        &...        &\\              
0933 & 1.48 & -0.2  &-1.2\tablenotemark{a} &-0.4 &-0.1    &-0.1       &$<$0.4  &...     &...        &1.1\tablenotemark{a}\\
1028 & 0.63 & $>$ 0.3  &$>$-0.4   &...  &$>$-0.2\tablenotemark{a} &... &... &$>$-0.6   &$>$-0.2\tablenotemark{a}  &0.1\tablenotemark{a}\\
1028 & 0.71 & $>$-0.5  &$>$-0.6 &... &$>$-0.8\tablenotemark{a} &... &... &$>$-0.2\tablenotemark{a} &$>$-0.3\tablenotemark{a} &-0.2\tablenotemark{a}\\
1107 & 0.74 & -0.3      &-0.4         &...      &...        &$<$-0.5 &$<-$0.1  &-0.6\tablenotemark{a}   &...        &...\\
1323 & 0.72 & -0.9      &-1.0         &...      &...        &...      &$<$0.1  &...        &...        &...\\
1727 & 0.95 & -0.4      &-0.7{a}     &-0.6     &-0.2       &-0.7 &$<-$0.3 &...        &...        &0.5\tablenotemark{a}\\
1727 & 1.03 & -0.3      &-1.0{a}     &-0.3     &0.0        &$<$0.0 &$<$0.1  &...        &-1.2       &1.0\tablenotemark{a} \\
2340 & 1.36 & -0.4      &-0.5         &$<$-0.5  &0.2        &$<$-0.2  &$<$0.4  &...        &-0.6       &0.7\\
\hline
Mean\tablenotemark{b}&$<1.5$ & -0.4      &-0.8         &-0.4     &0.0 &-0.4     &...     &-0.6       &-0.9       &0.5\\    
Mean\tablenotemark{c} &$>2.5$&-0.4 &-0.5         &-0.4     &0.1        &-0.9     &...     &...        &...        &0.4\\
\hline
GWC\tablenotemark{d,e} && -1.0     &-1.2         &-1.2     &-0.2       &-1.0     &-0.7    &-0.8       &-0.9       & 1.0\\
GCC\tablenotemark{d,f} && -1.7     &-1.8         &-1.8     &-0.9       &-2.4     &-1.7    &-1.1       &-2.0       &0.9\\
HC\tablenotemark{d,g}  && -0.5     &-0.5         &-0.5     &-0.2
&-0.6      &...     &-0.1       &...        &+0.3\\
\hline
\enddata

\tablenotetext{a}{Values based on data in Table 2, assuming $b$
value to be same as that of Zn}
\tablenotetext{b}{Note that the limits on relative abundances have been
excluded while calculating the mean values }
\tablenotetext{c}{Average values from Prochaska et al. (2003a)}
\tablenotetext{d}{Welty et al. (1997,1999a,1999b,2001), York et al.
(2005b)} 
\tablenotetext{e}{Galactic warm clouds } 
\tablenotetext{f}{Galactic cold clouds} 
\tablenotetext{g}{Galactic halo clouds} 
\end{deluxetable}
\clearpage
\setlength{\hoffset}{-0in}
\begin{table}
\centerline{Table 7 Reddening Characteristics for SDSS Quasars in Our Sample\label{tbl-7}}
\bigskip
\begin{tabular}{|l|l|l|l|l|l|l|}
\tableline
\multicolumn{1}{|l|}{\bf QSO}&\multicolumn{1}{|l|} {\bf $z_{em}$}&\multicolumn{1}{|l|}{\bf $\alpha_{RP}$}&\multicolumn{1}{|l|}{\bf
$E(B-V)_{RP}$}& \multicolumn{1}{|l|} 
{\bf $\Delta(g-i)$} & \multicolumn{1}{|l|} {\bf $E(B-V)_{g-i}$}&\multicolumn{1}{|l|}{\bf Visual Notes} \\
\tableline\tableline

SDSSJ1028-0100 & 1.531 &  ...   &     ...\tablenotemark{a}   &    -0.02 & ...&very blue \\
SDSSJ1107+0048 & 1.392 &  +0.30 &     0.05 &    0.14 & 0.05& very blue \\	
SDSSJ1323-0021 & 1.390 &  +0.06 &     0.17  &0.47 & 0.16&   somewhat red,\\
&&&&&&but no 2200 {\AA} dip \\
SDSSJ1727+5302 & 1.444 & +0.35 &     0.09 & 0.13 &0.04&   blue \\
SDSSJ2340-0053 & 2.085 &  -1.00 &     0.00  &0.39 & 0.09 &   somewhat red, \\
&&&&&&but no 2200 {\AA} dip \\
\tableline
\end{tabular}
\tablenotetext{a}{Reliable $E(B-V)_{RP}$ is not available for this object.}
\end{table}

\begin{table}
\centerline{Table 8 Rough Constraints on H I column densities and
abundances of Zn and Ti\label{tbl-8} }
\bigskip
\begin{tabular}{|l|l|l|r|r|}
\tableline
\multicolumn{1}{|l|}{\bf QSO}&\multicolumn{1}{|l|}{\bf $z_{abs}$}
&\multicolumn{1}{|l|}{\bf N$_{\rm H I}$}&\multicolumn{1}{|r|}{\bf
[Zn/H]}&\multicolumn{1}{|r|}{\bf [Ti/H]}\\
\tableline\tableline
Q0738+313     & 0.091  & 1.5$\times{10^{21}}^a$ &     &-1.6 \\
Q0738+313     & 0.221  & 7.9$\times{10^{20}}^a$ &     &$<$-1.7 \\
Q0933+733     & 1.479  & 4.2$\times{10^{21}}^a$ &-1.6 &-1.7\\
SDSSJ1107+0048 & 0.741 & 9.7$\times{10^{20}}^b$&-0.6&$<-1.1$\\
&                      &2.3$\times{10^{21}}^c$ &&\\
SDSSJ1323-0021 & 0.716 &1.6$ \times{10^{20}}^b$& 0.4&\\
&                      &7.5$\times{10^{21}}^c$ & & \\
SDSSJ1727+5302 & 0.945  & 1.5$\times{10^{21}}^d$ &-0.5 &-1.2 \\
SDSSJ1727+5302 & 1.031  & 2.6$\times{10^{21}}^d$ &-1.4 &$<$-1.4 \\
SDSSJ2340-0053 & 1.361  & 4.3$\times{10^{21}}^c$ &-1.6 & \\
\tableline
\end{tabular}
\tablenotetext{a:}{HST data: Rao \& Turnshek, (2000)}  
\tablenotetext{b:}{HST GO Project No. 9382, P.I. Rao, S.; see Table 5 for
error estimates}  
\tablenotetext{c:}{Estimates based on reddening}
\tablenotetext{d:}{HST data: Turnshek et al. (2004)}
\end{table}
\clearpage







\end{document}